\documentclass[traditabstract]{aa}

\usepackage{graphicx}   
\usepackage{txfonts}
\usepackage{amstext}
\usepackage[dvipsnames]{xcolor}
\usepackage{hyperref}
\hypersetup{
 colorlinks   = true, 
 urlcolor     = blue, 
 linkcolor    = blue, 
 citecolor    = blue  
 }

\newcommand{\m}{$^{\rm m}\!\!.$}

\newcommand{\ks}{km~s$^{-1}$}
\newcommand{\ms}{M$_{\odot}$}
\newcommand{\rs}{R$_{\odot}$}
\newcommand{\oc}{$O\!-\!C$ }
\newcommand{\hip}{{\sc Hipparcos}}
\newcommand{\T}{TESS}

\begin{document}

\title{Apsidal motion and TESS light curves of three southern close eccentric eclipsing binaries: GM~Nor, V397~Pup, and PT~Vel
\thanks{Based on observations secured at the 
        South Africa Astronomical Observatory, Sutherland, South Africa,
          Mt.~John University Observatory, University of Canterbury, Lake Tekapo, 
        New Zealand, 
        SPACE Atacama Lodge, San Pedro de Atacama, Chile, and 
        FRAM, Pierre Auger Observatory, Argentina.
    Tables~A1 -- A5, and B1, and B2 are available in the CDS via anonymous ftp to \url{cdsarc.cds.unistra.fr} (130.79.128.5) or via \url{https://cdsarc.cds.unistra.fr/viz-bin/cat/J/A+A/???/A??} \\
        \dag\, Martin Lehk\'y passed away on November~18, 2020.} }


\author{M. Wolf~\inst{1}
        \and P. Zasche~\inst{1}
        \and J. K\'{a}ra \inst{1}
        \and M. Zejda \inst{2} 
        \and J. Janík \inst{2}
        \and M. Ma\v{s}ek \inst{3, 4}
        \and M. Lehk\'y \inst{4, 5, \dag}
        \and J. Merc \inst{1} \and \\
             A. Richterkov\'a  \inst{2} 
        \and D. Han\v{z}l \inst{4,6}
        \and Z. Mikul\'a\v{s}ek \inst{2}
        \and S.N. de Villiers\inst{7}
        \and J. Li\v{s}ka \inst{2} 
       }
   \institute{Astronomical Institute, Faculty of Mathematics and Physics,
   Charles University Prague, V Hole\v{s}ovi\v{c}k\'ach 2, \\
   CZ-180 00 Praha 8, Czech Republic, E-mail: {\tt marek.wolf@mff.cuni.cz}   
    \and Institute of Theoretical Physics and Astrophysics, Masaryk
    University, Kotl\'a\v{r}sk\'a 2, CZ-611 37 Brno, Czech Republic 
  \and FZU - Institute of Physics of the Czech Academy of Sciences, Na Slovance 1999/2, 
  CZ-182 21, Praha 8, Czech Republic
      \and Czech Astronomical Society, Variable Star and Exoplanet Section, V\'ide\v{n}sk\'a~1056, CZ-142~00~Praha~4, Czech Republic
    \and Observatory and planetarium Hradec Kr\'alov\'e, Z\'ame\v{c}ek 456,
        CZ-500~08 Hradec Kr\'alov\'e, Czech Republic  
    \and Private Observatory, CZ-683~41 Pavlovice 65, Czech Republic 
    \and Private Observatory, Plumstead, Cape Town, South Africa 
   }

\date{Accepted XXX. Received YYY; in original form ZZZ}




\abstract{The study of apsidal motion in eccentric eclipsing binaries provides
an important observational test of theoretical models of stellar
structure and evolution. 
New ground-based and space-based photometric data have been obtained and archival spectroscopic measurements were used in this study of three detached early-type and southern-hemisphere eccentric eclipsing binaries 
   GM~Nor   ($P = 1\fd88, e = 0.05$),
   V397~Pup ($3\fd00, 0.30$), and
   PT~Vel   ($1\fd80, 0.12$). 
Their \T\ observations in several sectors have also been included and the corresponding light curves were solved using the {\sc Phoebe} code. 
As a result, new accurate photoelectric times of minimum light have been obtained.
The newly completed \oc\ diagrams were analyzed using all reliable timings found in the literature and calculated using the \T\ light curves.  New or improved values for the elements of apsidal motion were obtained. Using ESO archive spectroscopy, for V397~Pup, the precise absolute parameters were newly derived:  
$M_1$ =   3.076(35)~\ms,  
$M_2$ =   2.306(35)~\ms, and
$R_1$ =   2.711(55)~\rs,  
$R_2$ =   1.680(55)~\rs.
For PT~Vel the absolute dimensions were improved:
$M_1$ =   2.204(25)~\ms, 
$M_2$ =   1.638(25)~\ms, and
$R_1$ =   2.108(30)~\rs,  
$R_2$ =   1.605(30)~\rs.
For GM~Nor, the less accurate absolute parameters based on the light curve analysis were evaluated:
$M_1$ =   1.94(15)~\ms, 
$M_2$ =   1.84(14)~\ms, and
$R_1$ =   2.27(20)~\rs,  
$R_2$ =   2.25(20)~\rs.
We found more precise and relatively short periods of apsidal motion of about
80, 335, and 160 years, along with the corresponding internal structure constants, log $k_2$, 
--2.524, --2.361, and --2.563, for GM~Nor, V397~Pup, and PT~Vel, respectively.
Relativistic effects are small but not negligible, making up to 10\%\  of the total apsidal motion rate in all systems. 
No marks of the presence of the third body were revealed in the light curves, on the \oc diagrams, or in the reduced spectra of the eccentric systems studied here. }

\keywords{
binaries: eclipsing --
binaries: close --
stars: early-type --
stars: fundamental parameters --
stars: individual: GM~Nor, V397~Pup, PT~Vel }

\authorrunning{M. Wolf at al.}
\titlerunning{Apsidal motion and \T\ light curves of GM~Nor, V397~Pup, and PT~Vel}

\maketitle

\section{Introduction}

Eccentric eclipsing binaries (EEBs) provide an ideal opportunity for studying the stellar structure of the stars as well as testing general relativity effects outside of the Solar System.  
The Newtonian contribution to the observed apsidal motion in such
systems is highly dependent on stellar radii ($r^{-5}$). Therefore, only
systems with accurate determinations of their absolute dimensions
and precise apsidal motion periods can be used to test current stellar structure models
\citep{2010A&A...519A..57C, 2019ApJ...876..134C}.
A precise determination of the apsidal motion rate requires long-term monitoring
of mid-eclipse times, usually spanning several decades, and many advanced amateur
observers can help with this difficult task. 
Moreover, detached double-lined eclipsing binaries (DLEBs) simultaneously serve an important source of fundamental information on stellar masses and radii
\citep{1991A&ARv...3...91A}. 
\cite{2010A&ARv..18...67T}  revised the absolute dimensions of 94~DLEBs and only 18 of them are found to have eccentric orbits and show apsidal motion suitable for the apsidal motion test of the stellar structure.


 An analysis of apsidal motion in eclipsing binaries using \T\ data was presented by \cite{2021A&A...649A..64B, 2022A&A...665A..13B}. 
These authors determined the apsidal motion rate for nine EEBs, measured the general relativistic apsidal motion rate, and carried out a test of general relativity.
They found a perfect agreement with theoretical predictions and established stringent constraints on the parameters of the post-Newtonian formalism.
Recently, \cite{2021A&A...654A..17C} studied the internal structure constant
in 34~EEBs and nearly doubled the sample of suitable systems collected by \cite{2010A&ARv..18...67T}. Their comparison of the apsidal motion rate with predictions based on new theoretical models shows a very good agreement. 
Clearly, it is necessary to expand the current collection of EEBs with precise absolute parameters to provide better statistics on these results.

 Here, we analyze the available observational data and apsidal motion rates for three detached eclipsing systems. All these systems are relatively neglected bright southern-hemisphere objects of a similar spectral type that are 
known to have eccentric orbits and to exhibit relatively rapid apsidal motion. 
Their orbital periods are between 2 and 3 days. 
With the exception of PT~Vel, no radial velocities have been published for any of these binary systems. 
V397~Pup and PT~Vel were also included in the new Catalog of \T\ eclipsing binary stars collected by \cite{2022ApJS..258...16P}. 
Their basic properties are summarized in Table~\ref{summ}.
The previous \oc\ diagrams for these systems can also be found in the \oc\ gateway\footnote{\url{http://var.astro.cz/ocgate/}}.

This study of apsidal motion in EEBs is a continuation
of our work presented in earlier papers over several years
\citep{2008MNRAS.388.1836W, 2019AcA....69...63W, 2022NewA...9201708W}.
The current paper is organized as follows. 
Section~2 deals with new observations and data reductions. 
The apsidal motion analysis is given in Sect.~3, 
while the light curves and the radial velocity curves are analyzed in Sect.~4. 
The internal structure constants are derived in Sect.~5 
and a brief summary of the results presented is given in Sect.~6. 

\begin{table}
\caption {Basic properties of selected eclipsing binaries.
 The spectral type, $V$ magnitude, and parallax values in the last column 
 are taken from the {\sc Simbad} database. }
\label{summ}
\begin{tabular}{cccccc}
\hline\hline\noalign{\smallskip}
System   & Other      &  Spec. & Period  & $V$   & Par.   \\
         & name       &  type  &  [day]  & [mag] & [mas]  \\
\hline\noalign{\smallskip}
GM Nor   & CD-54~6427 &  A3V &  1.885 &  10.7  & 1.51   \\
V397 Pup & HD~63786   &  B9V &  3.004 &  5.94  & 6.70   \\
PT Vel   & HD~79154   &  A0V &  1.802 &  7.03  & 5.84   \\
\hline
\end{tabular}
\end{table}


The detached eclipsing binary GM~Nor 
(also CD $-54^{\circ} 6427$, CPD $-54^{\circ}~6718$, FL~1816, TIC~81569380;
$V_{\rm max}$ = 10\m6, Sp.~A3V) 
is a relatively poorly studied binary with a known and relatively fast apsidal motion. 
It was discovered to be a variable star by \cite{1932BAN.....6..233K} and later 
\cite{1946BAN....10..153P} derived the first light elements; 
\cite{1951AJ.....56....1S} included GM~Nor in the list of eclipsing binaries
with the displayed secondary eclipse.
\cite{1975A&AS...22..263S} obtained precise \textit{UBV} photometry using
the 50-cm telescope at La Silla, Chile, and found an apsidal motion with a period of about $90 \pm 15$ years.
The following light ephemeris were presented in the last-mentioned paper:

\begin{center}
Pri. Min. = HJD 24 41696$\fd$91 + 1$\fd$884577 $\cdot E$.  \\
\end{center}

\noindent 
Subsequently, GM~Nor was included in several catalogs of eccentric orbit eclipsing binaries \citep{2006A&A...446..785M, 2007MNRAS.378..179B, 2013AN....334..860A, 2018ApJS..235...41K}. 
For GM~Nor, the angular diameter of the system $\rho = 0.047$~mas was also measured by the {\it Gaia} satellite and this value is included
in the Mid-infrared stellar Diameters and Fluxes compilation Catalogue \cite [MDFC ver.~10] {2019MNRAS.490.3158C}.   

To our knowledge, no radial velocity data have been published for this
interesting and relatively bright system and almost 50 years have elapsed since the last published photometric study from \cite{1975A&AS...22..263S}. 
Surprisingly, no study that is directly focused on GM~Nor has been found in the literature. 
Furthermore, there is no spectrum of GM~Nor in the ESO Archive obtained in the past.


The detached eclipsing binary V397~Pup
(also HD~63786, HIP~38167, CD $-34^{\circ} 3970$, CPD $-34^{\circ} 1717$, TIC~150284425;
$V_{\rm max}$ = 5\m93, Sp.~B9V) 
is a relatively newly discovered bright early-type binary
with a significant eccentricity of $e \simeq 0.3$ and a short
orbital period of about 3.0 days.
It is a known visual binary system that was first suspected to be
a variable star (BV~438) by \cite{1964IBVS...66....1S}.
V397~Pup was discovered to be an EEB by \cite{2005IBVS.5631....1O} using 
the \hip\ and ASAS-3 photometry database. 
Otero also reported a relatively fast apsidal motion and derived the
following light ephemeris: 

\begin{center}
Pri. Min. = HJD 24 48799$\fd$646 + 3$\fd$004449 $\cdot E$,  \\
Sec. Min. = HJD 24 48801$\fd$672 + 3$\fd$004390 $\cdot E$.  \\
\end{center}

\noindent
With a period in days that is almost precisely an integer number,
V397~Pup is a rather difficult object to study from only one observatory.
Recently, V397~Pup was included in a new catalog of pulsating stars detected
in EA-type eclipsing binaries based on \T\ data \citep{2022ApJS..259...50S}.
To our knowledge, precise radial-velocity observations of this system
have not been published so far. We used 28 spectrograms of V397 Pup found in the ESO Archive and obtained in different projects during 2009 -- 2021, see Table~B1.
The \T\ light curve is also solved in the TESS eclipsing binary
catalog~\footnote{\url{http://tessebs.villanova.edu/}}, where an effective temperature, $T_{\rm eff} = 10\, 666$~K, was given. 

The first photometric observations of V397~Pup were made by the \hip\ mission 
and 138 points were collected, covering an interval of 3.2 years.
Unfortunately, only a few data points covering the eclipse
phases are available.  This is particularly true for the secondary eclipse phase.


The detached eclipsing binary PT~Velorum
(also HD~79154, CD$-42^{\circ}$~5038, HIP~45079, BV~469, TIC~74528791, NSV~04409;
$V_{\rm max}$ = 7\m03; Sp. A0V)
was a relatively little known early-type binary with a moderate
eccentric orbit ($e = 0.13$) and a short orbital period of about 1.8~days. 
It was discovered to be a variable by \cite{1964IBVS...66....1S} 
on photographic plates taken in Johannesburg.
Its eclipsing nature was revealed by \hip\ \citep{1997ESASP1200.....E}
and  \cite{2003IBVS.5482....1O} later found that the system exhibits an apsidal motion. Using high-resolution echelle spectra,
\cite{2008MNRAS.384.1657B} 
derived the absolute dimensions of the components and presented the
following linear light elements:

\begin{center}
Pri. Min. = HJD 24 48293$\fd$493 + 1$\fd$8020075 $\cdot E$, \\
Sec. Min. = HJD 24 48294$\fd$360 + 1$\fd$8020350 $\cdot E$.
\end{center}

\noindent
These authors confirmed a significant orbital eccentricity of $e = 0.127 \pm 0.006$
and a short period of apsidal motion of $U = 182.2 \pm 8.4$~year.
We also refer to the series of cited works published on this binary described
in that paper.
Using the \T\ data, \cite{2021A&A...654A..17C} recalculated the apsidal motion and the internal stellar structure of the system. 
The \T\ light curve is also solved in the TESS eclipsing binary
catalog, where the effective temperature $T_{\rm eff} = 10\, 239$~K is given. 
Recently, PT~Vel was included in the {\sc Banana} project \citep{2022ApJ...933..227M}, the study of spin-orbit alignment in close binaries.

\section{Observations}
\subsection{Ground-based photometry}

\begin{table}
\caption {Selected local comparison and check stars at different observatories.}
\label{comp}
\begin{tabular}{clcl}
\hline\hline\noalign{\smallskip}
Variable & Comparison \& & $V$   & Used for   \\
         & Check stars & [mag] &            \\
\hline\noalign{\smallskip}
GM Nor    & HD~141926*    & 9.13  &   PEP at SAAO  \\
          & GSC~8701.0957 & 10.4  &   CCD at SPA, Cape  \\
          &               &       &   Town and FRAM  \\
\hline\noalign{\smallskip}
V397 Pup  & HD~62781      & 5.78  &   PEP at SAAO  \\
          & HD~62578      & 5.59  &   PEP at SAAO  \\
          & HD~63077      & 5.36  &   CCD at SAAO  \\
          & HD~64379      & 5.01  &   CCD at SAAO  \\
\hline\noalign{\smallskip}
PT Vel  & HD~78429~$\dag$ & 7.33  &   PEP at MJUO  \\
          & HD 79735      & 5.25  &   CCD at SAAO  \\
          & HD 79810      & 6.80  &   CCD at SAAO  \\
          & HD 79415      & 8.50  &   CCD at SAAO  \\
\hline\noalign{\smallskip}
\end{tabular}

Notes: * HD~141926 used also in \cite{1975A&AS...22..263S}, \\
$\dag$ HD~78429 = Cousins E~475 southern standard star. 
\end{table}

Our new photoelectric observations were obtained at several different
southern observatories in the past, in chronological order: 

\begin{description}

\item{$\bullet$} Mt. John University Observatory (MJUO), 
University of Canterbury, Lake Tekapo, New Zealand:
0.61-m Cassegrain reflector (f/16) equipped with a
single-channel photon-counting photometer (utilizing an EMI 9202B 
photomultiplier) and Johnson \textit{UBV} filters or a SBIG ST-9 CCD camera with 
\textit{BVR} filters; covers over two weeks in January 2007. 

\item{$\bullet$} South African Astronomical Observatory (hereafer SAAO), 
Sutherland, South Africa: the 0.50-m Cassegrain reflector (f/18) equipped
with a modular photometer utilizing a Hamamatsu EA1516 photomultiplier and
Johnson \textit{UBV} filters or an objective Helios 2/58 with an ATIK 16 IC CCD
camera; during two weeks in November 2008 and April 2010. The objective Helios 2/58 with a camera G2-402 (\textit{BVR} filters) was used in March 2018. 

\item{$\bullet$} SPACE Atacama Lodge, San Pedro de Atacama (SPA), 
Chile: 0.20-m reflector (f/4) with a SBIG ST-8 CCD camera and \textit{VR} filters; during two weeks in April 2010. 

\item{$\bullet$} Fotometric Robotic Atmospheric Monitor \citep[FRAM]{2014RMxAC..45..114E},
Pierre Auger Observatory, Argentina: 0.30-m reflector with a G2-1600 CCD camera and \textit{BVRI} filters; during two epochs in October 2013 and June 2014.

\item{$\bullet$} Private observatory of S.N. de Villiers in Plumstead, Cape Town, South Africa: Celestron 280/1764 telescope, a SBIG ST-8XME CCD camera, and \textit{BVRI} filters during 13 nights in the 2019--2020 season.   

\end{description}

\noindent
The main aim of these photometric measurements was to obtain the whole light curve
or at least to secure several well-covered primary and secondary eclipses
for all variables. 
 Due to the limited observation time, complete light curves were obtained only for GM Nor.
Each observation of an eclipsing binary was accompanied by the observation
of a local comparison and check stars (see Table~\ref{comp}).  
We note that in the neighborhood of PT~Vel there is also a bright $\delta$-Sct variable star MP~Vel ($V = 7.8$ mag) which is not suitable as a comparison.

Photoelectric measurements at MJUO and SAAO were usually performed using Johnson's \textit{UBV} photometric filters with a 10-second integration time.  
All observations were carefully reduced to the Cousins E~region standard
system \citep{1989SAAOC..13....1M} and corrected for differential extinction
using the reduction program HEC~22 rel.~16 \citep{1998JAD.....4....5H}.
The standard errors of the \textit{UBV} measurements at SAAO were approximately 0.009,
0.007 and 0.006 mag in the $U, B$ and $V$ filters, respectively. The software package
{\sc C-Munipack}\footnote{Motl, 2018, \url{https://c-munipack.sourceforge.net/}, ver. 2.1.24} 
for aperture photometry, based on the {\sc Daophot} procedure, was
used to process the set of CCD frames. 
Examples of our photometric observations of GM~Nor obtained at SPA and Cape Town observatories are given in Fig.~\ref{gmv}.

\subsection{\T\ photometry}

Moreover, the Transiting Exoplanet Survey Satellite (TESS) mission
to study exoplanets through photometric transits \citep{2015JATIS...1a4003R}, 
with its nearly full sky coverage, also provides precise photometry of a large
sample of eclipsing binary systems with a time baseline of 27 days to several years. 
Thus, precise monitoring of light curves is possible from space and their exceptional
quality is perfect for studying light curves and eclipse timings \citep{2021A&A...649A..64B}. 
The three systems studied in this work have been observed by \T\ in several sectors (see Table~\ref{tess}). 
 For GM Nor and PT Vel, we collected simple aperture photometry (SAP) of available cadence produced by the Science Process Operation Centre (SPOC) \citep{2016SPIE.9913E..3EJ} available at Mikulski Archive for Space Telescope
(MAST)\footnote{\url{https://mast.stsci.edu/portal/Mashup/Clients/Mast/Portal.html }}. 
In the case of V397~Pup, the original FFI data from the \T\ archive were downloaded and reduced using the {\tt Lightkurve} code\footnote{\url{https://github.com/lightkurve/lightkurve}}
in the standard procedure.

\begin{table}
\caption {\T\ visibility and sectors used for light curve analyses and for mid-eclipse time determination.}
\label{tess}
\begin{tabular}{cccccc}
\hline\hline\noalign{\smallskip}
System   &  TIC        & Sector &  Start & Exposure    \\
         &  number     &   No.  &  date  &  time [sec]  \\
\hline\noalign{\smallskip}
GM Nor   & 81569380  &    12      & 2019-05-21 & 1800  \\
         &           &    39      & 2021-05-27 &  600  \\
         &           &    65      & 2023-05-04 &  158  \\ 
\hline\noalign{\smallskip}
V397 Pup & 150284425 &     7      & 2019-01-08 &  120   \\
         &           &     8      & 2019-02-02 &  120    \\
         &           &    34      & 2021-01-14 &  120    \\
         &           &    61      & 2023-01-18 &  120    \\   
\hline\noalign{\smallskip}
PT Vel   & 74528791  &     8      & 2019-02-02 &  1800   \\
         &           &     9      & 2019-03-25 &  1800   \\
         &           &    35      & 2021-02-09 &  600    \\
         &           &    36      & 2021-03-07 &  600    \\
         &           &    62      & 2023-02-12 &  200    \\
\hline
\end{tabular}
\end{table}

The new times of primary and secondary minima and their errors
were determined using a least-squares fit of the light curve during eclipses. Three or fourth-order polynomial fittings were applied to the bottom sections of the light curves. The mean values of the individual filter bands are given and the presented errors represent the fitting mean error for each light curve. 
For the \T\ data, the mid-eclipse times were determined by
{\sc Silicups}\footnote{SImple LIght CUrve Processing System, \\
\url{https://www.gxccd.com/cat?id=187&lang=405}} code.

Because the \T\ data are provided in the Barycentric Julian Date Dynamical Time  (BJD$_{\rm TDB}$), all our previous times were first transformed to this time
scale using the often used Time Utilities of Ohio State 
University\footnote{\url{http://astroutils.astronomy.ohio-state.edu/time/} } 
\citep{2010PASP..122..935E}.

\subsection{Hipparcos, ASAS-3, and OMC photometry}

Using the \hip\ photometry \citep{1997ESASP1200.....E}, taken from the Epoch Photometry Annex, 
All Sky Automated Survey-3 (ASAS-3) 
database\footnote{\url{http://www.astrouw.edu.pl/asas/}} \citep{2002AcA....52..397P}, 
and photometric data from the {\sc OMC Integral} Optical
Monitoring Camera Archive 
\citep{2004ESASP.552..729M}\footnote{\url{https://sdc.cab.inta-csic.es/omc/}} we were able to derive
several additional times of minimum light with a lower precision. 
They were used in our next analysis with lower weight.

To derive some of the eclipse times from various photometric surveys, we used the following approach. Due to having only sparse photometry (i.e., typically a few data points per night only), 
we used our method named AFP (described in our paper \cite{2014A&A...572A..71Z}). It uses the phased light curve over a longer time interval and the template of the light curve provided through light curve modeling in {\sc Phoebe}.
The time interval used was typically one year, but this can be arbitrarily changed with respect to the number of data points in each interval. We are typically able to derive several times of eclipses, primary and secondary, using this method from one particular photometric survey dataset.

All new minima times are included in Tables~A1 -- A5. 
In these tables, the epochs are calculated from the light ephemeris given in the text.  The other columns are self-evident.

\begin{figure}
\centering
\includegraphics[width=0.47\textwidth]{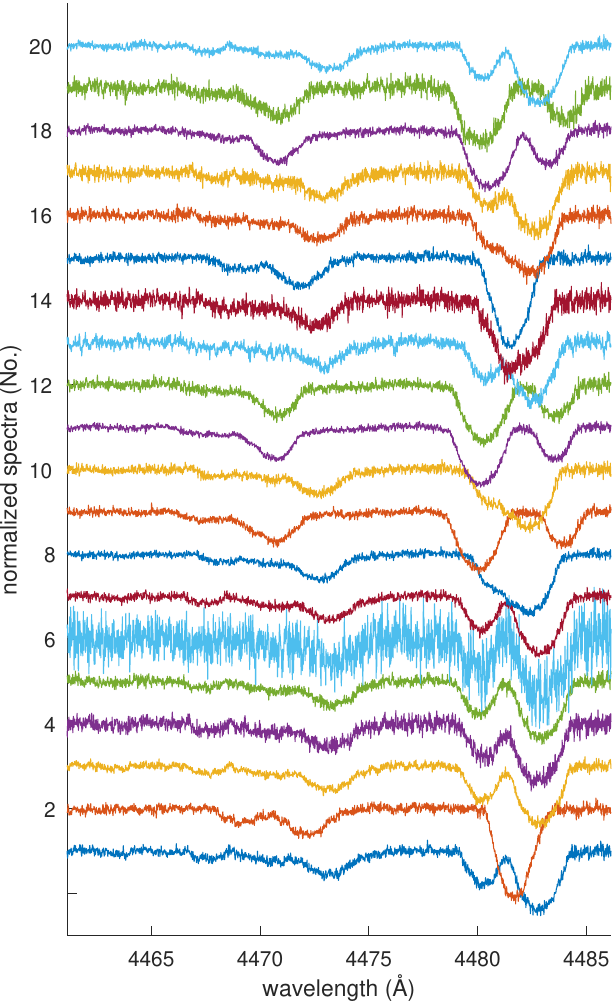}
\caption{Sample of 20 HARPS spectra of V397~Pup with prominent 
\ion{He}{4471} and \ion{Mg}{4481} absorption lines.}
\label{397harps}
\end{figure}

\subsection{Spectroscopy}

The only system in this study with previously published spectroscopy and a radial velocity curve is PT~Vel \citep{2008MNRAS.384.1657B}. A total of 28 radial velocities obtained using the {\sc Hercules} spectrograph at Mount John University Observatory, Lake Tekapo, New Zealand, in May 2006, are available.

The ESO HARPS -- High Accuracy Radial velocity Planet 
Searcher\footnote{\url{https://www.eso.org/sci/facilities/lasilla/instruments/harps.html}}  \citep{2003Msngr.114...20M} 
and FEROS -- Fiber-fed Extended Range Optical
Spectrograph\footnote{\url{https://www.eso.org/public/teles-instr/lasilla/mpg22/feros/}}
have been used in several projects for spectroscopy of our objects in the past\footnote{082.D-0499 PI Pietrzynski, 086.D-0078 PI Gieren, 091.D-0414 PI Graczyk}. 
We retrieved the publicly available HARPS spectroscopic data sets from the ESO Science data archive\footnote{\url{https://archive.eso.org/eso/eso_archive_main.html}}.
For V397~Pup we found 28 original spectra in the Archive, for PT~Vel there were
additional 16 spectra at our disposal.
All spectra were reduced in the standard way using IRAF\footnote{\label{iraf}IRAF is distributed by NOAO, which is operated by AURA, Inc., under cooperative agreement with the National Science Foundation.} routines. The spectra were bias-corrected and flat-fielded. From the archival spectra, the absorption lines of prominent Balmer lines of H$\alpha$, H$\beta$, or H$\gamma$ were detected. Unfortunately, the broad hydrogen lines exceeded the aperture size, so the more suitable \ion{Mg}{4481} and \ion{Fe}{5018} lines were used to measure the radial velocities (see Fig.~\ref{397harps} and Tables~B1 and B2). The Gaussian fit with the IRAF program was used to determine individual radial velocities.
To our knowledge, for a relatively bright target GM~Nor, no spectroscopic material or radial velocities have been reported in the literature or found in any electronic database.


\section{Apsidal motion analysis}
\label{apsmotion}

Precise determination of the apsidal motion requires long-term
monitoring of eclipse times, usually spanning several decades. 
As usual, the apsidal motion in EEBs can be studied using an 
\oc\ diagram (ETV curve) analysis. 
As in our previous studies, the iterative method described by \cite{1983Ap&SS..92..203G} or \cite{1995Ap&SS.226...99G} was used routinely.
We collected all reliable times of minimum light available in
the literature and current databases. The input files of the mid-eclipse times
were supplemented with our own measurements and numerous data from \T.
We computed all \T\ mid-eclipse times from the whole available light curves.  
All modern photoelectric and CCD times, as well as \T\ minima were used with a weight of 10 in our calculation. Previous, less accurate measurements with unknown uncertainty (photographic or visual estimations) were assigned weights of 3, 1, or 0.

\begin{table*}
\centering
\caption{Apsidal motion elements of GM~Nor, V397~Pup, and PT~Vel. }
\label{t2}
\begin{tabular}{clllll}
\hline\hline\noalign{\smallskip}
Element      &  Unit  &  GM~Nor & V397~Pup & PT~Vel  \\
\hline\noalign{\smallskip}
$T_0$        & BJD      & 41683.7157 (5)   & 48502.4531 (6) & 48293.4745 (8) \\
$P_s$        & days     & 1.88457721 (12)  & 3.0044198 (7)  & 1.8020229 (6)  \\
$P_a$        & days     & 1.88469921 (12)  & 3.0044936 (8)  & 1.8020782 (6)  \\
$e$          &  --      & 0.0478 (12)      & 0.285 (3)      & 0.1195 (2) \\
$\dot{\omega}_{\rm obs}$ & deg $\:\rm{cycle^{-1}}$ 
                                    & 0.0233 (14) & 0.0088 (4)  & 0.01105 (8) \\
$\dot{\omega}_{\rm obs}$ & deg $\:\rm{yr^{-1}}$  
                                    & 4.52 (27)   & 1.07 (7)  & 2.24 (13) \\
$\omega_0$   & deg                  & 178.8 (1.1) & 19.1 (1.0) & 257.5 (0.8) \\
$U$          & years                & 79.7 (4.8)  & 335 (25)  & 160.7 (13) \\ \hline\noalign{\smallskip}
$\Sigma w (O-C)^2$ & day$^2$ 
             & $4.1\cdot10^{-3}$ & $9.23\cdot10^{-3}$ & $1.93\cdot10^{-3}$ \\ 
\hline
\end{tabular}
\end{table*}

Apsidal motion parameters were calculated iteratively. For the orbital inclination, we adopted the results from our own photometric solution
(see Chapter~\ref{lcrv} and Table~\ref{gm}). For the eccentricity, we adopted the value from the apsidal motion analysis. In general, the apsidal motion solution was found to be insensitive to relatively large changes in inclination but strongly depends on the orbital eccentricity.


\begin{figure}
\centering
\includegraphics[width=0.47\textwidth]{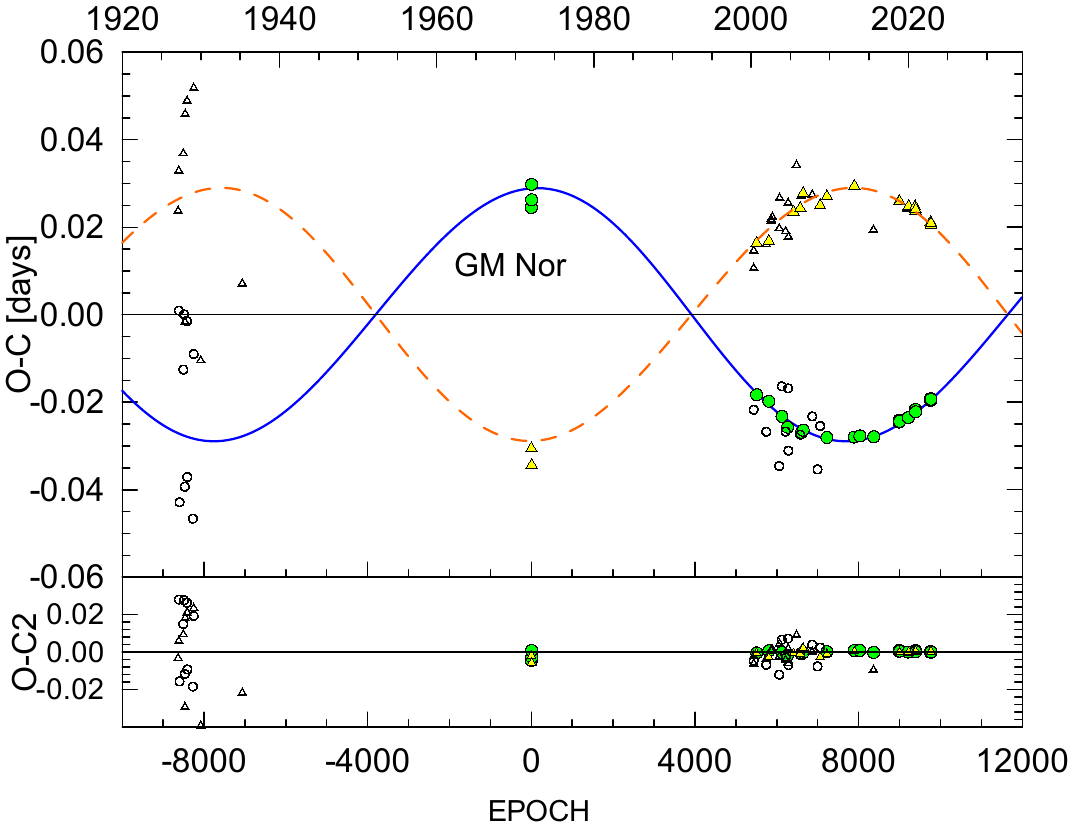}
\caption[ ]{Historical \oc\ diagram of GM~Nor together with our best-fit apsidal motion model.
          The blue solid curve and green circles correspond to the primary, 
          the orange curve and yellow triangles to the secondary minima.
          Larger symbols denote the photoelectric or CCD measurements
          which were given higher weights in the calculations.
          The $O-C_2$ residuals are plotted in the bottom panel.} 
\label{gmnor}
\end{figure}

\begin{figure}
\centering
\includegraphics[width=0.40\textwidth]{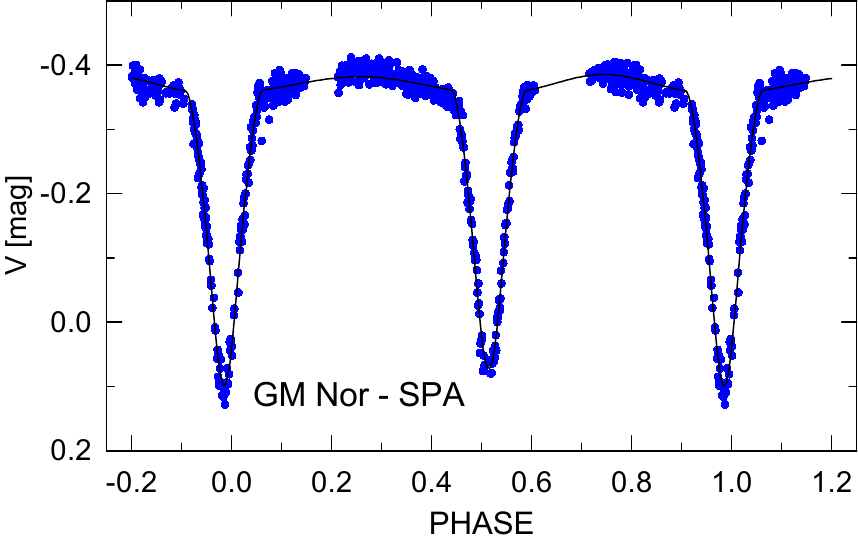}

\medskip
\includegraphics[width=0.4\textwidth]{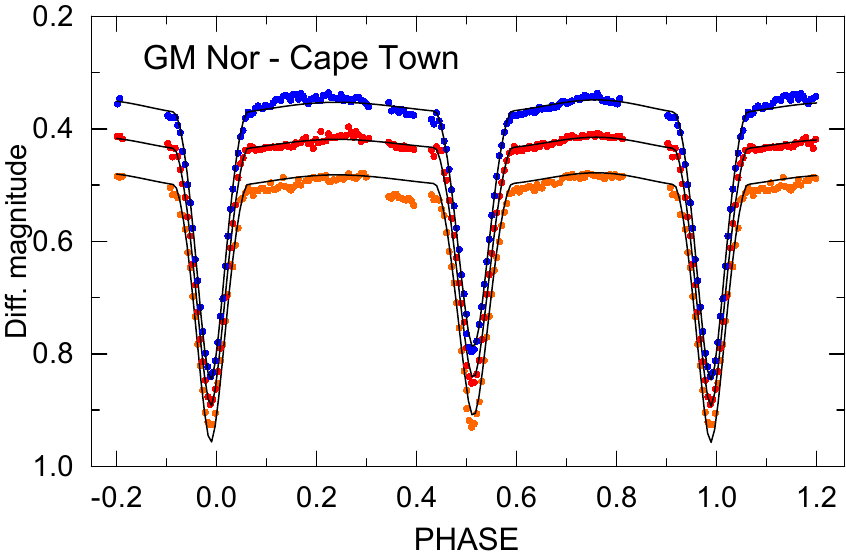}
\caption{Examples of our observations. Upper panel:  Observed differential $V$ light curve of GM~Nor obtained at SPA observatory (blue dots) in April 2010. 
Lower panel: Observed $VRI$ light curves of GM~Nor obtained at Cape Town in 2020. The {\sc Phoebe} solutions are plotted as black curves.}
\label{gmv}
\end{figure}

 In case of GM Nor, all photoelectric times of minimum light given in \cite{1946BAN....10..153P} and \cite [his Table~7] {1975A&AS...22..263S} were incorporated into our calculation.
Our new $VR$ CCD photometry obtained at the SPA (1590 measurements in both filters) and Cape Town observatories (see Fig.~\ref{gmv}) was used for the determination of new mid-eclipse times. 
Several times with less precision were derived using ASAS and OMC photometry. 
A large number of very precise and consecutive times were obtained using
the \T\ data in three sectors. A total of 165 times of minimum light were used in our analysis, with 82 secondary eclipses among them. All are listed in Tables~A1 and~A2.

The calculated parameters of the apsidal motion and their internal
errors of the least-squares fit (in brackets) are given in Table~\ref{t2}.
In this table, $P_s$ denotes the sidereal period, $P_a$ the anomalistic
period, $e$ represents the eccentricity, and
$\dot{\omega}_{\rm obs}$ is the rate of advance of the periastron
(in degrees per cycle or in degrees per year).
The zero epoch is given by $T_0$, and the corresponding position of
the periastron is represented by $\omega_0$.
The orbital eccentricity was taken as a free parameter in our
calculations and was used for the light-curve solution (see Chapter~\ref{lcrv}). The resulting values of $e$ have lower intrinsic
errors compared to those determined independently from the light curve
analysis. This procedure usually gives us a better result for this important element. The \oc\ diagram is shown in Fig.~\ref{gmnor}.

\begin{table*}
\centering
\caption{Temperatures of selected binaries found in different sources.}
\label{temp1}  
\begin{tabular}{ccccc}
\hline\hline\noalign{\smallskip}
 Source                  & GM Nor &   V397 Pup  &  PT Vel & Reference  \\
\hline\noalign{\smallskip}
Modern Mean Dwarf Stellar Color &  A3V  &  B9V  & A0V & {\sc Simbad} \\
and Effective Temperature Sequence & 8 600 & 10 700 &  9 700 & \cite{2013ApJS..208....9P}   \\
    VOSA 7.5             & 7 500  & 10 000  &  9250   &  \cite{2008AA...492..277B} \\
TESS Input Catalog & 8 398  & 10 666  & 10 239  &  \cite{2019AJ....158..138S}   \\      
{\it Gaia} DR3           & 8 078(20) & 10 552(32) & 8 712(4) &  \cite{2022yCat.1355....0G}   \\
\noalign{\smallskip}\hline
\end{tabular}
\end{table*}

\begin{figure}
\centering
\includegraphics[width=0.47\textwidth]{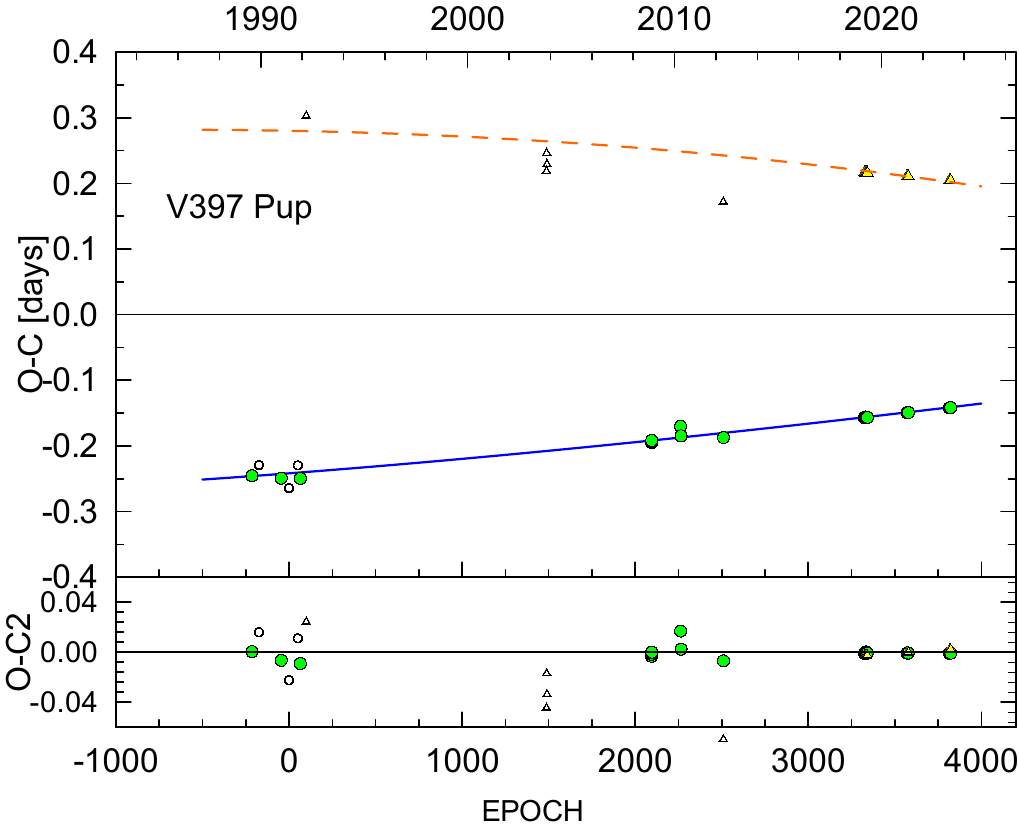}
\caption[ ]{ \oc\ diagram for V397~Pup.
            See legend to Fig.~\ref{gmnor}.
            Although primary minima fit the curve well, the secondary have a larger scatter. Minima derived from \T\ sectors are bulks of points close to epoch 3400, 3600 and 3800. }
\label{397pup}
\end{figure}

 For V397 Pup, all times of minimum light given in \cite [his Table~2]{2005IBVS.5631....1O}, 
our new timings derived from the \hip\ database,
the results of our new measurements at SAAO observatory,
as well as numerous \T\ eclipse times were included in our analysis.
The epochs in Table~A3 were calculated according
to the basic minimum given in the \hip\ catalog.
A total of 78 times of minimum light were used in our
analysis, with 35 secondary eclipses among them.
The calculated parameters of the apsidal motion and their internal errors
of least squares fit are given in Table~\ref{t2}. 
The current \oc\ diagram is shown in Fig.~\ref{397pup}.


\begin{figure}
\includegraphics[width=0.47\textwidth]{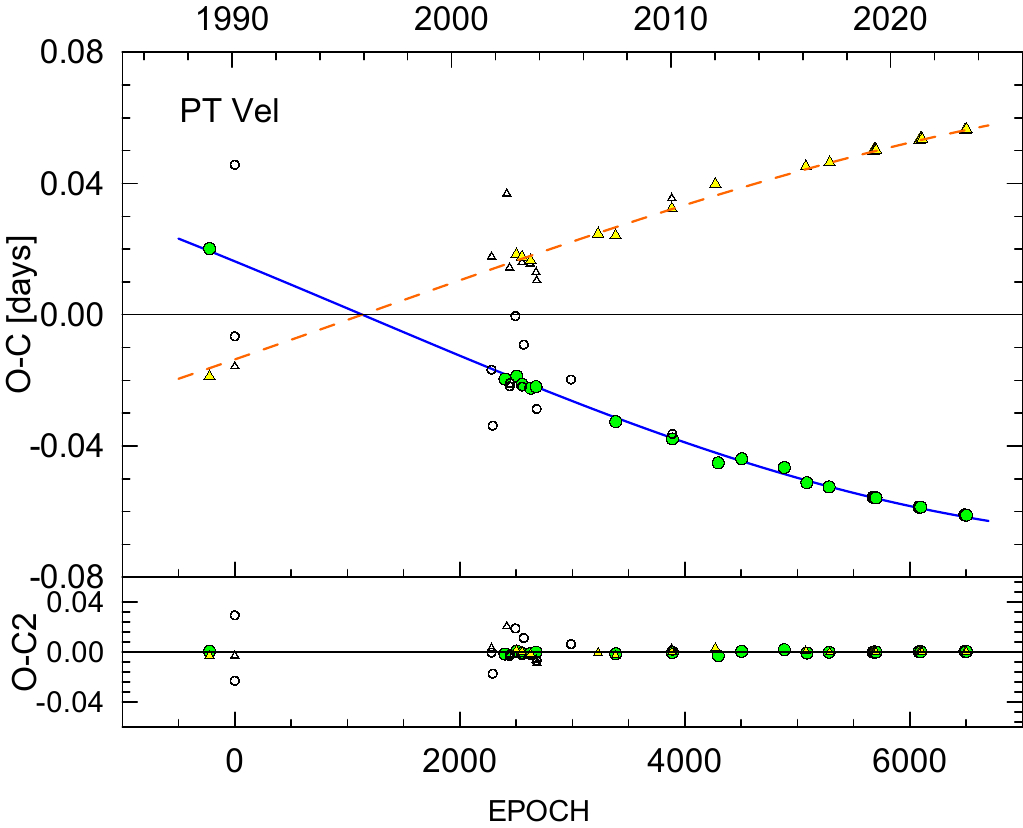}
\caption{ Current \oc\ diagram for PT~Vel. See legend to Fig.~\ref{gmnor}. Numerous minima derived from \T\ sectors are groups of points close to epoch 5700, 6100, and 6500.}
\label{ptvel}
\end{figure}

 In case of PT~Vel, all mid-eclipse times given in \cite[his Table~2]{2003IBVS.5482....1O} were incorporated into our analysis. 
Seven new and precise times were derived using our SAAO and MJUO observations.
Additional timings derived from the HIP photometry (135~points spanning an interval of about three~years), ASAS and {\sc Integral OMC} camera were included. 
The usefulness of the \hip\ data is partially limited by the lack of observations during the secondary minimum.  The fully covered and rather scattered $V$ light curve is available in the ASAS-3 Photometric Catalogue.
The most numerous contribution of minima comes from five \T\ sectors. 
The newly determined times of minimum light are listed
in Tables~A4 and A5, where all epochs were computed according to the ephemeris given in \cite{2008MNRAS.384.1657B}.
In our analysis, a total of 154 times of minimum light were used. 
The residuals of \oc\ values for all minimum times with respect to the
linear part of the apsidal motion equation are shown in Fig.~\ref{ptvel}.
The computed elements of apsidal motion and their internal errors
of least squares fit are given in Table~\ref{t2}. 
The nonlinear predictions, corresponding to the
fitted parameters, are plotted as blue and orange curves for
primary and secondary eclipses, respectively.

\section{Light curve and radial velocities analysis}
\label{lcrv}

\begin{table}
\caption{Parameters of the fit to the \T\ light curve for GM~Nor, V397~Pup and PT~Vel.}
\label{gm}
\centering
\begin{tabular}{lccc}
\hline\hline\noalign{\smallskip}
 Element           & GM Nor      & V397 Pup      & PT Vel \\
\hline\noalign{\smallskip}
 $T_1$ [K] (fixed) & 8\,500(200) &  10\,500(200) &  9\,250(200) \\
 $T_2$ [K]         & 8\,320(200) &  9\,300(200)  &  7\,690(200) \\
 $r_1 = R_1/a$     &  0.227(8)   &  0.178        &  0.216 \\
 $r_2 = R_2/a$     &  0.225(8)   &  0.107        &  0.165 \\
 $e$ (fixed)       & 0.0479      &  0.286        &  0.120 \\ 
 $i$ [deg]         & 83.5(0.5)   & 78.9(0.8)     & 88.5(0.4) \\
 $\Omega_1$        & 3.81(3)   & 4.43(2)    &   3.69(1)   \\
 $\Omega_2$        & 3.15(3)   & 2.90(2)    &   2.90(1)   \\
 \hline\noalign{\smallskip}
 $\chi^2_{\rm red}$ & 3.14    &  7.56   &   5.22   \\
\noalign{\smallskip}\hline
\end{tabular}
\end{table}

\begin{figure}
\centering
\includegraphics[width=0.45\textwidth]{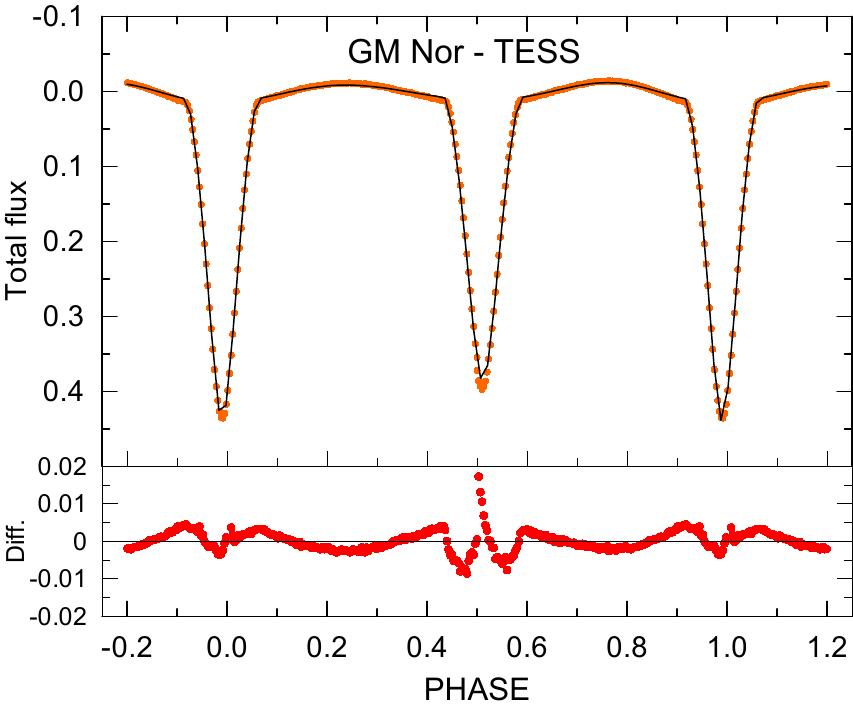}
\caption{\T\ light curve of GM~Nor obtained in Sector~12 (May/June 2019, orange circles) and its {\sc Phoebe} solution (black curve). 
The residuals from the solution are plotted in the bottom panel.}
\label{gmtess0}
\end{figure}

\begin{figure}
\centering
\includegraphics[width=0.45\textwidth]{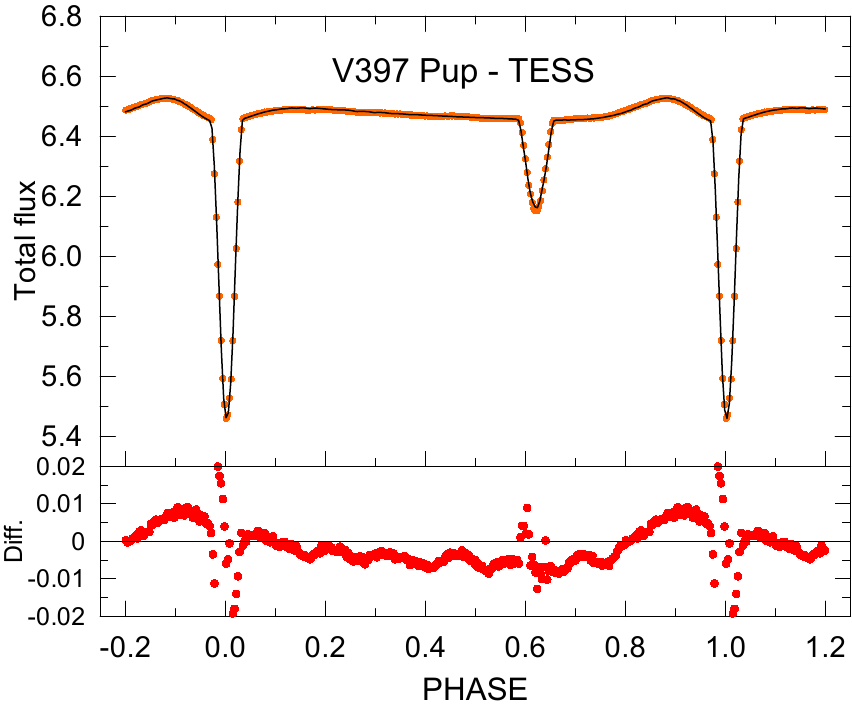}
\medskip
\includegraphics[width=0.45\textwidth]{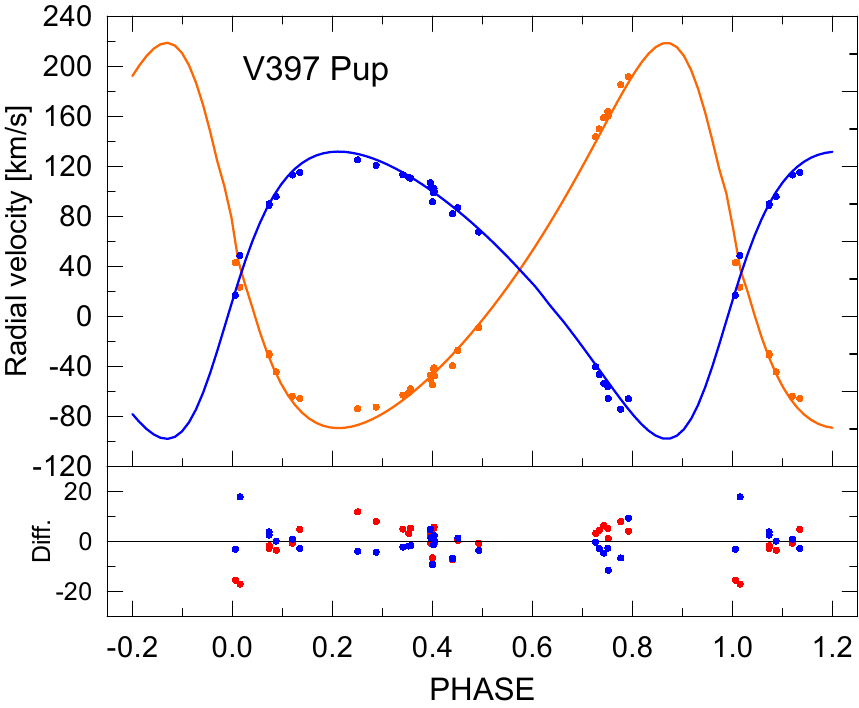}
\caption{Solution of light and radial velocity curves for V397~Pup. 
         Upper panel: \T\ light curve (orange circles, binning 300) 
         and its {\sc Phoebe} solution (black curve).
         Lower panel:  Radial velocity curve of V397~Pup  
and its {\sc Phoebe} solution (primary component in orange, secondary in blue). 
The residuals of the solution are plotted in the bottom panels.}
\label{397lcfinal}
\end{figure}

\begin{figure}
\centering
\includegraphics[width=0.45\textwidth]{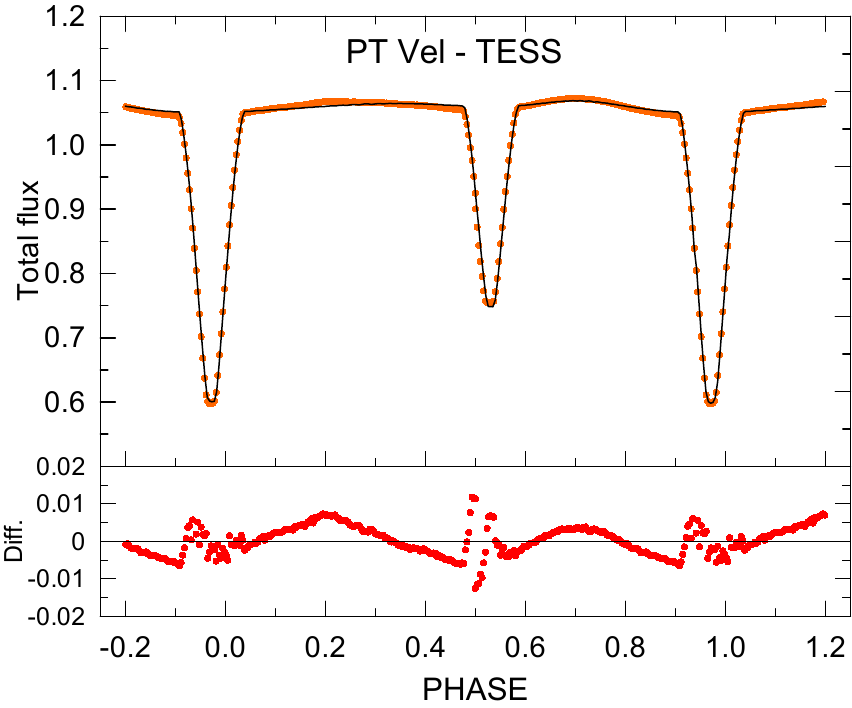}
\medskip
\includegraphics[width=0.45\textwidth]{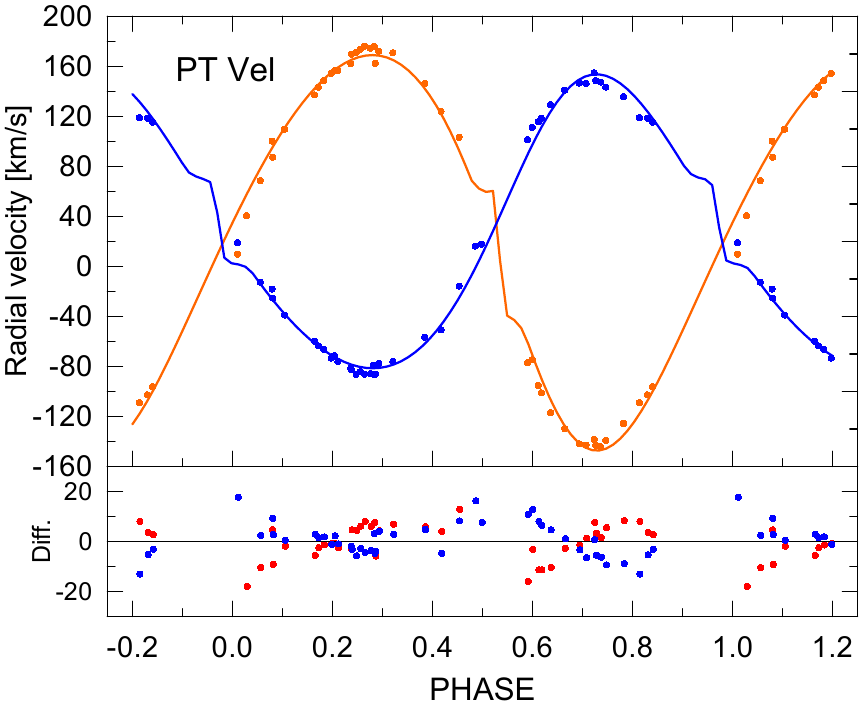}
\caption{Solution of light and radial velocity curves for PT~Vel. 
         Upper panel: \T\ light curve 
        (orange circles, binning 300) and its {\sc Phoebe} solution (black curve).
Lower panel:  Radial velocity curve of PT~Vel obtained by 
\cite{2008MNRAS.384.1657B} and new data, and its {\sc Phoebe} solution. 
Primary component in blue, secondary component in orange.
 The residuals of the solution are plotted in the bottom panels.}
\label{ptlcfinal}
\end{figure}

Due to the relatively lower quality data obtained at the MJUO, SAAO, SPA, FRAM, and Cape Town observatories compared to the \T\ data, only the \T\ light curves were selected to fit the light curves of three systems. 
The completely covered light curves were routinely analyzed using the
{\sc Phoebe} computer code, version 31c, developed by \cite{2005ApJ...628..426P}, see also \cite{2018maeb.book.....P}, 
which is a user-friendly implementation of the well-known Wilson-Devinney code
\citep{1971ApJ...166..605W}. 
To reduce the long computation time for eccentric binaries, 
the \T\ light curves were phased-binned to 300 points each.
The other photometric data available to us were used mostly for mid-eclipse time determination and solution of apsidal motion. 

The light elements and the apsidal motion parameters (eccentricity, 
position of the periastron, and apsidal motion rate) have been adopted
from the apsidal motion analysis (see Sect.~\ref{apsmotion} and Table~\ref{t2}) 
because the minima times cover a longer time span and these quantities are
derived with higher precision.
The other light curve parameters have been fitted: the luminosities,
 temperature of the secondary,  inclination, and Kopal's
modified potentials.
The temperature of the primary component was adopted 
as a mean value of several sources or fixed in agreement with previous investigators (see Table~\ref{temp1}).
 
The fitting of the $F_1, F_2$ synchronicity parameters in {\sc Phoebe} was unstable. 
During the procedure, their resulting values oscillated between 0.8 -- 2.0.
Thus, we assumed the synchronous rotation for both components ($F_1 = F_2 = 1$).
Moreover, in the case of elliptical orbits, the rotation tends to synchronize because of tidal interactions between the two components.
The limb-darkening coefficients were interpolated from van Hamme's
tables \citep{1993AJ....106.2096V}, using the linear cosine law.
The fine and coarse grid rasters for both components were set at 30.
Numerous {\sc Phoebe} runs were evaluated in detached mode using a different
configuration of initial parameters. The results and the value of the cost function were recorded.
The final solution was accepted when subsequent iterations did not result in a decrease of the {\sc Phoebe} cost function.

The results of the photometric analysis are given in Table~\ref{gm}. 
In this table, $T_1, T_2$ denote the temperatures of the primary and secondary
components, $r_1, r_2$ the relative radii, and $i$ the orbital inclination. 
The corresponding light curves are plotted in Fig.~\ref{gmtess0}, \ref{397lcfinal} and \ref{ptlcfinal}, respectively.
The \T\ phased-binned light curve of V397~Pup shows asymmetric maxima (Fig.~\ref{397lcfinal}). These out-of-eclipse variations are caused by the effect of light reflection from one component to the other and are well resolved using the {\sc Phoebe} code.

The results of the radial velocity analysis and absolute dimensions
are listed in Table~\ref{t3}.  For GM~Nor, the dimensions were estimated
using the {\sc Phoebe} solution and the spectral type and were interpolated in the table of \cite{2013ApJS..208....9P}\footnote{A Modern Mean Dwarf Stellar Color and Effective Temperature Sequence \url{https://www.pas.rochester.edu/~emamajek/EEM_dwarf_UBVIJHK_colors_Teff.txt}}.


\begin{figure}[t]
\centering
\includegraphics[width=0.45\textwidth]{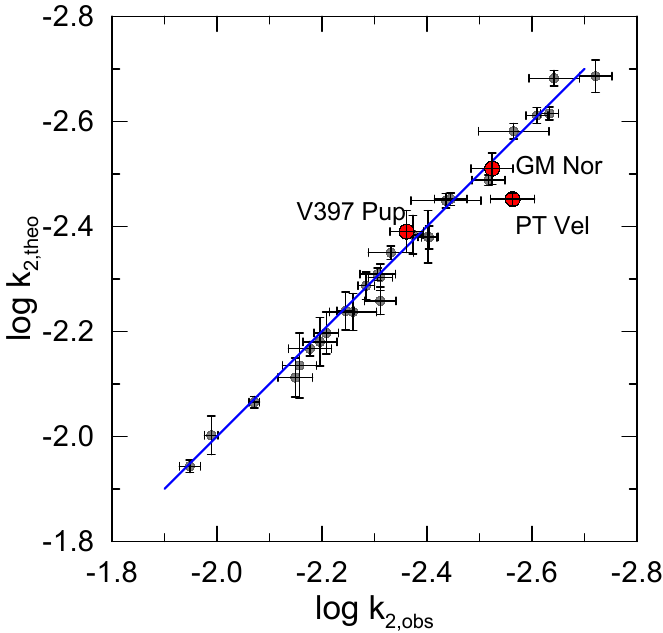}
\caption{Comparison of theoretical and observed internal structure 
constants of GM~Nor, V397~Pup and PT~Vel (red dots) with other systems collected in \cite [their Table 3] {2021A&A...654A..17C}.}
\label{k2comp}
\end{figure}

\begin{table*}
\centering
\caption{Astrophysical parameters and internal structure constants.} 
\label{t3}
\begin{tabular}{lccccccc}
\hline\hline\noalign{\smallskip}
Parameter &  Unit &  GM~Nor* & V397~Pup & PT~Vel \\
\hline\noalign{\smallskip}
$M_1$   & \ms  &  1.94(0.15)  & 3.076(35)  &  2.204(25)  \\
$M_2$   & \ms  &  1.84(0.14)  & 2.306(35)  &  1.638(25)  \\
$R_1$   & \rs  &  2.27(0.20)   & 2.711(55)  &  2.108(30) \\
$R_2$   & \rs  &  2.25(0.20)   & 1.680(55)  &  1.605(30) \\
$L_1$   &  L\sun &  22.4(2.5)  & 74.3(8.7)   &  27.0(3.1)   \\
$L_2$   &  L\sun &  20.2(2.4)  & 17.6(2.7)   &  7.5(1.0)    \\
$\log g_1$ & cgs &  4.0(0.6)  & 4.059(20)  &  4.133(18)   \\
$\log g_2$ & cgs &  3.9(0.5)  & 4.335(20)  &  4.241(18)   \\
$q = M_2/M_1$ & -- & 0.95  & 0.750      &  0.743   \\
$a$      & \rs  &  10.0(fixed)  & 15.35(15)  &  9.76(8)  \\
$\gamma$ & \ks  &  --     & 36.8(2.5)  &  25.6(1.8)  \\  
\hline\noalign{\smallskip}
$\dot{\omega}_\mathrm{rel}$  & deg $\:\rm{cycle^{-1}}$ 
                              & 0.000\ 87 & 0.000\ 88 & 0.000\ 92 \\
$\dot{\omega}_\mathrm{rel} / \dot{\omega}$ & \% &  3.7 & 9.7 &  8.3 \\
Age         &  year      &  $7\cdot10^8$ & $3\cdot10^8$  & $5\cdot10^8$  \\  
log $k_\mathrm{2, obs}$  & -- &  --2.524(40) & --2.361(32)  &  --2.563(42) \\
log $k_\mathrm{2, theo}$ & -- &  --2.51(3)   & --2.38(4)    &  --2.452(12)**  \\
\hline
\end{tabular}

\medskip
Note: * {\sc Phoebe} LC estimation, ** value taken from \cite{2021A&A...654A..17C}
\end{table*}

\begin{figure}[h]
\centering
\includegraphics[width=0.45\textwidth]{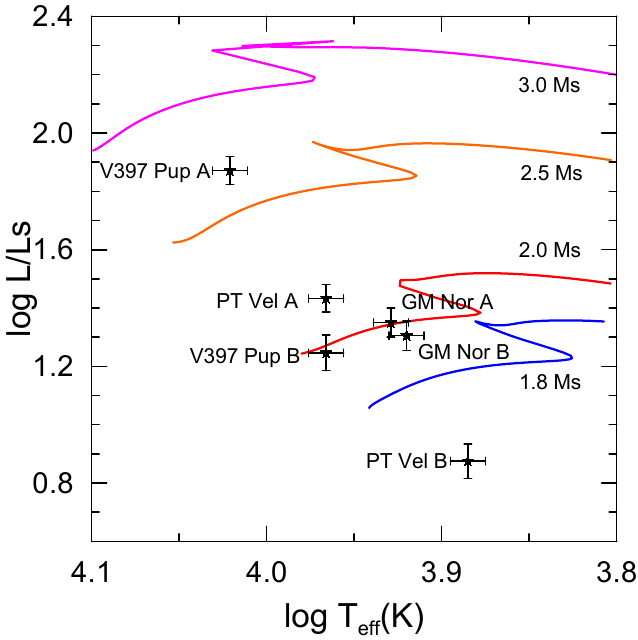}
\includegraphics[width=0.45\textwidth]{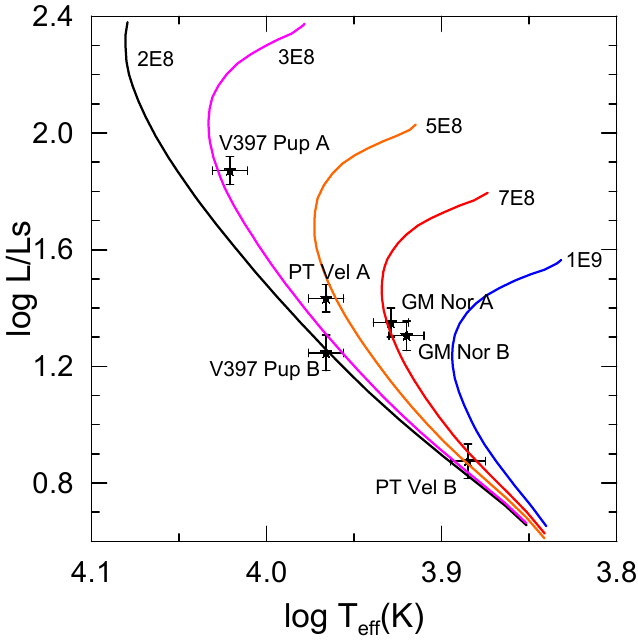}
\caption[]{HR~diagram for components of studied binaries. 
          Upper panel: Models of stellar evolution according to \cite{2019A&A...628A..29C} 
          for masses of 1.8, 2.0, 2.5, and 3.0 \ms\ are plotted. 
          Bottom panel: Isochrones calculated for ages from $2\cdot 10^8$ 
          to $10^9$ yr according to CMD 3.7 and PARSEC 2.0 models are given.}
\label{HRD}
\end{figure}


\section{Internal structure constant}

Observations of binary systems allow us to determine the internal
structure constant (ISC), $k_2$, which is a measure of the density
variation within the star and is an important parameter of stellar
evolution models. It is best studied in binary systems with eccentric
orbits that show apsidal motion.
The mean value of observed $\overline k_{2,\rm obs}$ is given by

\begin{equation}
\overline k_\mathrm{2, obs} = \frac{1}{c_{21} + c_{22}} \, \frac{P_a}{U}
         = \frac{1}{c_{21} + c_{22}} \, \frac{\dot{\omega}}{360} ,
\end{equation}

\noindent
where $c_{21}$ and $c_{22}$ are functions of the orbital eccentricity,
fractional radii,  masses of the components, and the ratio between
rotational velocity of the stars and Keplerian velocity \citep{1978ASSL...68.....K}.
The rotation of the stars was assumed to be synchronized with
the maximum angular orbital velocity achieved at periastron.
The astrophysical parameters of all systems are collected in Table~\ref{t3}.

The observed apsidal motion rate has two independent components: the Newtonian (classical) term, due to the non-spherical shape of both stars, and the relativistic term due to general relativity effects. 
Taking into account the value of the eccentricity and the masses of the components, 
a relativistic term, $\dot{\omega}_{\rm rel}$, must first be subtracted.
The original equation by \cite{1937AJMat..59..225L}  could be rewritten to be a more
suitable function of known observable parameters, in degrees per cycle
\citep{1985ApJ...297..405G}:

\begin{equation}
\dot{\omega}_\mathrm{rel} = \; 5.447 \times 10^{-4} \: \frac{1}{1-e^2}
          \:\Biggl( \frac{M_1 + M_2}{P} \Biggr) ^{2/3},
\end{equation}
\smallskip

\noindent where $M_i$ denotes the individual masses of the
components in solar units and $P$ is the orbital period in days.
The value of the classical contribution to the observed
apsidal motion rate, $\dot{\omega}_\mathrm{cl}$ is then the following:

\begin{equation}
\dot{\omega}_\mathrm{cl} = \dot{\omega}_\mathrm{obs}  - \dot{\omega}_\mathrm{rel}.
\end{equation}
\smallskip

\noindent
The values of $\dot{\omega}_\mathrm{rel}$ and the resulting mean internal
structure constants $\overline k_\mathrm{2,obs}$ are given in Table~\ref{t3}.
Their errors were determined using the relation derived in
\cite{2005AA...437..545W}. 

For PT~Vel we used for comparison the theoretical values of the ISC, $k_{\rm 2, theo}$  from updated models according to the masses adopted and the chemical composition of \cite{2021A&A...654A..17C}. 
These models assume convective core overshooting following the semi-empirical dependence on mass given by \cite{2018ApJ...859..100C}.
For GM~Nor and V397~Pup, we roughly estimated the theoretical value of $k_{\rm 2, theo}$ from the tables of \cite [Tables 14-17] {2023A&A...674A..67C} interpolating the given mass, expected age of the components (see Table~\ref{t3} and conclusions below), and a mean chemical composition $Z = 0.0134$.
The comparison of the theoretical and observed internal structure constant
with previously studied binaries \citep [their Table 3] {2021A&A...654A..17C} is shown in Fig.~\ref{k2comp}.

\section{Conclusions}

This study provides accurate information on the apsidal motion rates, absolute parameters, and observed ISC values for the three southern detached binary systems GM~Nor, V397~Pup, and PT~Vel. All derived or confirmed periods of apsidal motion are relatively short.
The obtained values of the mean ISC $\overline k_{\rm 2,obs}$ are compared
to their theoretical values $k_{\rm 2,theo}$ according to the current theoretical
models along the main sequence computed by \cite{2023A&A...674A..67C}.
With the exception of PT~Vel, the agreement is very good (see Fig.~\ref{k2comp}).

Our results for apsidal motion and light-curve analyses of GM~Nor 
are similar to the previous parameters derived in \cite{1975A&AS...22..263S}. 
GM~Nor also presents the shortest period of apsidal motion in our small sample ($\sim$ 80~years). Most likely, the system consists of two similar stars of A3$\pm$1 spectral type. 
It was also confirmed that V397~Pup is an EEB illustrated by a significant
apsidal motion with a relatively short period of about 330 years.
The precise masses of both components $M_1$ = 3.08 \ms\ and $M_2$ = 2.31 \ms\ were newly derived. 
For PT~Vel, we found a slightly shorter apsidal motion period
of $U = 161 \pm 13$ years than that given in \cite{2008MNRAS.384.1657B}.
Due to the limited coverage of the apsidal motion period in V397~Pup and PT~Vel, additional observations over longer time periods are necessary to further constrain the results presented in this work.
In this respect, our results should be considered as preliminary, since the characterization of these systems would still benefit from further photometric and spectroscopic observations.

The new and improved absolute dimensions of the components of V397~Pup and PT~Vel are given in Table~\ref{t3}, with relative errors up to 3 \%. 
These values are sufficiently precise for comparison with current theoretical models. Their positions in the HR~diagram that contains the evolutionary models for different masses according to \cite{2019A&A...628A..29C} is plotted in Fig.~\ref{HRD}.  
In the bottom part of this figure, the position of the components is compared with the set of isochrones calculated for the ages from $2\cdot 10^8$ to $10^9$ years, according to the CMD 3.7 web interface and the PARSEC 2.0 models \citep{2012MNRAS.427..127B} available on the web pages of Osservatorio Astronomico di Padova.~\footnote{\url{http://stev.oapd.inaf.it/cgi-bin/cmd\_3.7}.  }
Both systems, V397~Pup and PT~Vel, are now part of an increasing group of eclipsing binaries with precise absolute dimensions that are suitable for subsequent tests.
In these systems, no indications of the presence of a third component were observed.
In the case of GM~Nor, we do not yet know the precise absolute parameters. It would also be desirable to obtain new high-dispersion and high-S/N spectroscopic observations for GM~Nor and to apply modern disentangling methods to obtain the radial velocity curve and derive accurate masses for this system. 

\medskip
\begin{acknowledgements}
Useful suggestions and recommendations from an anonymous referee helped improve
the clarity of the article and are greatly appreciated.
This investigation was supported
by allocation of SAAO, MJUO and SPA observation time.
Research of MW, PZ, JK and JM was partially supported by the project
{\sc Cooperatio - Physics} of the Charles University in Prague.
This publication was produced within the framework of institutional support 
for the development of the research organization of Masaryk University.
We wish to thank Alan C. Gilmore, Mt.~John University Observatory,
Dr.~Alain Maury, SPACE Atacama Lodge, and the staff at SAAO for their warm hospitality and help with the equipment.
The authors thank J.~Jury\v{s}ek and K.~Ho\v{n}kov\'a, Variable Star and Exoplanet Section,
for their important contribution to photometric observations with the FRAM telescope. 
This paper includes data collected by the \T\ mission. The funding for the \T\ mission is provided by the NASA Science Mission Directorate. 
Data presented in this paper were obtained from the Mikulski Archive for Space Telescopes (MAST).
We thank the Pierre Auger Collaboration for the use of its facilities. The operation of the FRAM robotic telescope was supported by the EU grant GLORIA (No. 283783 in FP7-Capacities program). This work is supported by MEYS (Czech Republic) under the projects MEYS LM2023032, CZ.02.1.01/0.0/0.0/16\_013/0001402, CZ.02.1.01/0.0/0.0/18\_046/0016010 and CZ.02.1.01/0.0/0.0/ 17\_049/0008422 and CZ.02.01.01/00/22\_008/0004632. 
This work is supported by grants GACR~24-13049S and CAS LQ100102401.
This publication makes use of VOSA, developed under the Spanish Virtual Observatory (https://svo.cab.inta-csic.es) project funded by MCIN/AEI/10.13039/501100011033/ through the grant PID2020-112949GB-I00.
VOSA has been partially updated using funding from the European Union's Horizon 2020 Research and Innovation Programme, under Grant Agreement No. 776403 (EXOPLANETS-A).
The following Internet-based resources were used in the research for this paper:
the SIMBAD database operated at CDS, Strasbourg, France,
the NASA's Astrophysics Data System Bibliographic Services, 
the OMC Archive at LAEFF, preprocessed by ISDC. 
This investigation is part of an ongoing collaboration between
professional astronomers and the Czech Astronomical Society, 
Variable Star and Exoplanet Section.
\end{acknowledgements}

\bibliographystyle{aa_link}
\bibliography{pceb} 

\begin{thebibliography}{55}
\expandafter\ifx\csname natexlab\endcsname\relax\def\natexlab#1{#1}\fi

\bibitem[{{Andersen}(1991)}]{1991A&ARv...3...91A}
{Andersen}, J. 1991, \href{http://dx.doi.org/10.1007/BF00873538}{\color{magenta}\aapr}, \href{https://ui.adsabs.harvard.edu/abs/1991A&ARv...3...91A}{3, 91}

\bibitem[{{Avvakumova} {et~al.}(2013){Avvakumova}, {Malkov}, \& {Kniazev}}]{2013AN....334..860A}
{Avvakumova}, E.~A., {Malkov}, O.~Y., \& {Kniazev}, A.~Y. 2013, \href{http://dx.doi.org/10.1002/asna.201311942}{\color{magenta}Astronomische Nachrichten}, \href{https://ui.adsabs.harvard.edu/abs/2013AN....334..860A}{334, 860}

\bibitem[{{Bak{\i}{\c{s}}} {et~al.}(2008){Bak{\i}{\c{s}}}, {Bak{\i}{\c{s}}}, {Demircan}, \& {Eker}}]{2008MNRAS.384.1657B}
{Bak{\i}{\c{s}}}, V., {Bak{\i}{\c{s}}}, H., {Demircan}, O., \& {Eker}, Z. 2008, \href{http://dx.doi.org/10.1111/j.1365-2966.2007.12822.x}{\color{magenta}\mnras}, \href{https://ui.adsabs.harvard.edu/abs/2008MNRAS.384.1657B}{384, 1657}

\bibitem[{{Baroch} {et~al.}(2022){Baroch}, {Gim{\'e}nez}, {Morales}, {Ribas}, {Herrero}, {Perdelwitz}, {Jordi}, {Granzer}, \& {Allende Prieto}}]{2022A&A...665A..13B}
{Baroch}, D., {Gim{\'e}nez}, A., {Morales}, J.~C., {et~al.} 2022, \href{http://dx.doi.org/10.1051/0004-6361/202244287}{\color{magenta}\aap}, \href{https://ui.adsabs.harvard.edu/abs/2022A&A...665A..13B}{665, A13}

\bibitem[{{Baroch} {et~al.}(2021){Baroch}, {Gim{\'e}nez}, {Ribas}, {Morales}, {Anglada-Escud{\'e}}, \& {Claret}}]{2021A&A...649A..64B}
{Baroch}, D., {Gim{\'e}nez}, A., {Ribas}, I., {et~al.} 2021, \href{http://dx.doi.org/10.1051/0004-6361/202040004}{\color{magenta}\aap}, \href{https://ui.adsabs.harvard.edu/abs/2021A&A...649A..64B}{649, A64}

\bibitem[{{Bayo} {et~al.}(2008){Bayo}, {Rodrigo}, {Barrado Y Navascu{\'e}s}, {Solano}, {Guti{\'e}rrez}, {Morales-Calder{\'o}n}, \& {Allard}}]{2008AA...492..277B}
{Bayo}, A., {Rodrigo}, C., {Barrado Y Navascu{\'e}s}, D., {et~al.} 2008, \href{http://dx.doi.org/10.1051/0004-6361:200810395}{\color{magenta}\aap}, \href{https://ui.adsabs.harvard.edu/abs/2008A&A...492..277B}{492, 277}

\bibitem[{{Bressan} {et~al.}(2012){Bressan}, {Marigo}, {Girardi}, {Salasnich}, {Dal Cero}, {Rubele}, \& {Nanni}}]{2012MNRAS.427..127B}
{Bressan}, A., {Marigo}, P., {Girardi}, L., {et~al.} 2012, \href{http://dx.doi.org/10.1111/j.1365-2966.2012.21948.x}{\color{magenta}\mnras}, \href{https://ui.adsabs.harvard.edu/abs/2012MNRAS.427..127B}{427, 127}

\bibitem[{{Bulut} \& {Demircan}(2007)}]{2007MNRAS.378..179B}
{Bulut}, {\.I}. \& {Demircan}, O. 2007, \href{http://dx.doi.org/10.1111/j.1365-2966.2007.11756.x}{\color{magenta}\mnras}, \href{https://ui.adsabs.harvard.edu/abs/2007MNRAS.378..179B}{378, 179}

\bibitem[{{Claret}(2019)}]{2019A&A...628A..29C}
{Claret}, A. 2019, \href{http://dx.doi.org/10.1051/0004-6361/201936007}{\color{magenta}\aap}, \href{https://ui.adsabs.harvard.edu/abs/2019A&A...628A..29C}{628, A29}

\bibitem[{{Claret}(2023)}]{2023A&A...674A..67C}
{Claret}, A. 2023, \href{http://dx.doi.org/10.1051/0004-6361/202346250}{\color{magenta}\aap}, \href{https://ui.adsabs.harvard.edu/abs/2023A&A...674A..67C}{674, A67}

\bibitem[{{Claret} \& {Gim{\'e}nez}(2010)}]{2010A&A...519A..57C}
{Claret}, A. \& {Gim{\'e}nez}, A. 2010, \href{http://dx.doi.org/10.1051/0004-6361/201014008}{\color{magenta}\aap}, \href{https://ui.adsabs.harvard.edu/abs/2010A&A...519A..57C}{519, A57}

\bibitem[{{Claret} {et~al.}(2021){Claret}, {Gim{\'e}nez}, {Baroch}, {Ribas}, {Morales}, \& {Anglada-Escud{\'e}}}]{2021A&A...654A..17C}
{Claret}, A., {Gim{\'e}nez}, A., {Baroch}, D., {et~al.} 2021, \href{http://dx.doi.org/10.1051/0004-6361/202141484}{\color{magenta}\aap}, \href{https://ui.adsabs.harvard.edu/abs/2021A&A...654A..17C}{654, A17}

\bibitem[{{Claret} \& {Torres}(2018)}]{2018ApJ...859..100C}
{Claret}, A. \& {Torres}, G. 2018, \href{http://dx.doi.org/10.3847/1538-4357/aabd35}{\color{magenta}\apj}, \href{https://ui.adsabs.harvard.edu/abs/2018ApJ...859..100C}{859, 100}

\bibitem[{{Claret} \& {Torres}(2019)}]{2019ApJ...876..134C}
{Claret}, A. \& {Torres}, G. 2019, \href{http://dx.doi.org/10.3847/1538-4357/ab1589}{\color{magenta}\apj}, \href{https://ui.adsabs.harvard.edu/abs/2019ApJ...876..134C}{876, 134}

\bibitem[{{Cruzal{\`e}bes} {et~al.}(2019){Cruzal{\`e}bes}, {Petrov}, {Robbe-Dubois}, {Varga}, {Burtscher}, {Allouche}, {Berio}, {Hofmann}, {Hron}, {Jaffe}, {Lagarde}, {Lopez}, {Matter}, {Meilland}, {Meisenheimer}, {Millour}, \& {Schertl}}]{2019MNRAS.490.3158C}
{Cruzal{\`e}bes}, P., {Petrov}, R.~G., {Robbe-Dubois}, S., {et~al.} 2019, \href{http://dx.doi.org/10.1093/mnras/stz2803}{\color{magenta}\mnras}, \href{https://ui.adsabs.harvard.edu/abs/2019MNRAS.490.3158C}{490, 3158}

\bibitem[{{Eastman} {et~al.}(2010){Eastman}, {Siverd}, \& {Gaudi}}]{2010PASP..122..935E}
{Eastman}, J., {Siverd}, R., \& {Gaudi}, B.~S. 2010, \href{http://dx.doi.org/10.1086/655938}{\color{magenta}\pasp}, \href{https://ui.adsabs.harvard.edu/abs/2010PASP..122..935E}{122, 935}

\bibitem[{{Ebr} {et~al.}(2014){Ebr}, {Jane{\v{c}}ek}, {Prouza}, {Kub{\'a}nek}, {Jel{\'\i}nek}, {Ma{\v{s}}ek}, {Ebrov{\'a}}, \& {{\v{C}}ern{\'y}}}]{2014RMxAC..45..114E}
{Ebr}, J., {Jane{\v{c}}ek}, P., {Prouza}, M., {et~al.} 2014, in Revista Mexicana de Astronomia y Astrofisica Conference Series, Vol.~45, Revista Mexicana de Astronomia y Astrofisica Conference Series, \href{https://ui.adsabs.harvard.edu/abs/2014RMxAC..45..114E}{114}

\bibitem[{{Gaia Collaboration}(2022)}]{2022yCat.1355....0G}
{Gaia Collaboration}. 2022, \href{https://ui.adsabs.harvard.edu/abs/2022yCat.1355....0G}{VizieR Online Data Catalog, I/355}

\bibitem[{{Gim\'{e}nez}(1985)}]{1985ApJ...297..405G}
{Gim\'{e}nez}, A. 1985, \href{http://dx.doi.org/10.1086/163539}{\color{magenta}\apj}, \href{https://ui.adsabs.harvard.edu/abs/1985ApJ...297..405G}{297, 405}

\bibitem[{{Gim{\'e}nez} \& {Bastero}(1995)}]{1995Ap&SS.226...99G}
{Gim{\'e}nez}, A. \& {Bastero}, M. 1995, \href{http://dx.doi.org/10.1007/BF00626903}{\color{magenta}\apss}, \href{https://ui.adsabs.harvard.edu/abs/1995Ap&SS.226...99G}{226, 99}

\bibitem[{{Gim\'{e}nez} \& {Garcia-Pelayo}(1983)}]{1983Ap&SS..92..203G}
{Gim\'{e}nez}, A. \& {Garcia-Pelayo}, J.~M. 1983, \href{http://dx.doi.org/10.1007/BF00653602}{\color{magenta}\apss}, \href{https://ui.adsabs.harvard.edu/abs/1983Ap&SS..92..203G}{92, 203}

\bibitem[{{Harmanec} \& {Horn}(1998)}]{1998JAD.....4....5H}
{Harmanec}, P. \& {Horn}, J. 1998, Journal of Astronomical Data, \href{https://ui.adsabs.harvard.edu/abs/1998JAD.....4....5H}{4, 5}

\bibitem[{{Jenkins} {et~al.}(2016){Jenkins}, {Twicken}, {McCauliff}, {Campbell}, {Sanderfer}, {Lung}, {Mansouri-Samani}, {Girouard}, {Tenenbaum}, {Klaus}, {Smith}, {Caldwell}, {Chacon}, {Henze}, {Heiges}, {Latham}, {Morgan}, {Swade}, {Rinehart}, \& {Vanderspek}}]{2016SPIE.9913E..3EJ}
{Jenkins}, J.~M., {Twicken}, J.~D., {McCauliff}, S., {et~al.} 2016, in Society of Photo-Optical Instrumentation Engineers (SPIE) Conference Series, Vol. 9913, Software and Cyberinfrastructure for Astronomy IV, ed. G.~{Chiozzi} \& J.~C. {Guzman}, \href{https://ui.adsabs.harvard.edu/abs/2016SPIE.9913E..3EJ}{99133E}

\bibitem[{{Kim} {et~al.}(2018){Kim}, {Kreiner}, {Zakrzewski}, {Og{\l}oza}, {Kim}, \& {Jeong}}]{2018ApJS..235...41K}
{Kim}, C.~H., {Kreiner}, J.~M., {Zakrzewski}, B., {et~al.} 2018, \href{http://dx.doi.org/10.3847/1538-4365/aab7ef}{\color{magenta}\apjs}, \href{https://ui.adsabs.harvard.edu/abs/2018ApJS..235...41K}{235, 41}

\bibitem[{{Kopal}(1978)}]{1978ASSL...68.....K}
{Kopal}, Z. 1978, {Dynamics of close binary systems}

\bibitem[{{Kruytbosch}(1932)}]{1932BAN.....6..233K}
{Kruytbosch}, W.~E. 1932, \bain, \href{https://ui.adsabs.harvard.edu/abs/1932BAN.....6..233K}{6, 233}

\bibitem[{{Levi-Civita}(1937)}]{1937AJMat..59..225L}
{Levi-Civita}, T. 1937, Am. Jour. Math, \href{https://doi.org/10.2307/2371404}{59, 225}

\bibitem[{{Malkov} {et~al.}(2006){Malkov}, {Oblak}, {Snegireva}, \& {Torra}}]{2006A&A...446..785M}
{Malkov}, O.~Y., {Oblak}, E., {Snegireva}, E.~A., \& {Torra}, J. 2006, \href{http://dx.doi.org/10.1051/0004-6361:20053137}{\color{magenta}\aap}, \href{https://ui.adsabs.harvard.edu/abs/2006A&A...446..785M}{446, 785}

\bibitem[{{Marcussen} \& {Albrecht}(2022)}]{2022ApJ...933..227M}
{Marcussen}, M.~L. \& {Albrecht}, S.~H. 2022, \href{http://dx.doi.org/10.3847/1538-4357/ac75c2}{\color{magenta}\apj}, \href{https://ui.adsabs.harvard.edu/abs/2022ApJ...933..227M}{933, 227}

\bibitem[{{Mas-Hesse} {et~al.}(2004){Mas-Hesse}, {Gim{\'e}nez}, {Domingo}, {R{\'\i}squez}, {Caballero}, {Guti{\'e}rrez}, {Solano}, \& {OMC Team}}]{2004ESASP.552..729M}
{Mas-Hesse}, J.~M., {Gim{\'e}nez}, A., {Domingo}, A., {et~al.} 2004, in ESA Special Publication, Vol. 552, 5th INTEGRAL Workshop on the INTEGRAL Universe, ed. V.~{Schoenfelder}, G.~{Lichti}, \& C.~{Winkler}, \href{https://ui.adsabs.harvard.edu/abs/2004ESASP.552..729M}{729}

\bibitem[{{Mayor} {et~al.}(2003){Mayor}, {Pepe}, {Queloz}, {Bouchy}, {Rupprecht}, {Lo Curto}, {Avila}, {Benz}, {Bertaux}, {Bonfils}, {Dall}, {Dekker}, {Delabre}, {Eckert}, {Fleury}, {Gilliotte}, {Gojak}, {Guzman}, {Kohler}, {Lizon}, {Longinotti}, {Lovis}, {Megevand}, {Pasquini}, {Reyes}, {Sivan}, {Sosnowska}, {Soto}, {Udry}, {van Kesteren}, {Weber}, \& {Weilenmann}}]{2003Msngr.114...20M}
{Mayor}, M., {Pepe}, F., {Queloz}, D., {et~al.} 2003, The Messenger, \href{https://ui.adsabs.harvard.edu/abs/2003Msngr.114...20M}{114, 20}

\bibitem[{{Menzies} {et~al.}(1989){Menzies}, {Cousins}, {Banfield}, \& {Laing}}]{1989SAAOC..13....1M}
{Menzies}, J.~W., {Cousins}, A.~W.~J., {Banfield}, R.~M., \& {Laing}, J.~D. 1989, South African Astronomical Observatory Circular, \href{https://ui.adsabs.harvard.edu/abs/1989SAAOC..13....1M}{13, 1}

\bibitem[{{Otero}(2003)}]{2003IBVS.5482....1O}
{Otero}, S.~A. 2003, Information Bulletin on Variable Stars, \href{https://ui.adsabs.harvard.edu/abs/2003IBVS.5482....1O}{5482, 1}

\bibitem[{{Otero}(2005)}]{2005IBVS.5631....1O}
{Otero}, S.~A. 2005, Information Bulletin on Variable Stars, \href{https://ui.adsabs.harvard.edu/abs/2005IBVS.5631....1O}{5631, 1}

\bibitem[{{Pecaut} \& {Mamajek}(2013)}]{2013ApJS..208....9P}
{Pecaut}, M.~J. \& {Mamajek}, E.~E. 2013, \href{http://dx.doi.org/10.1088/0067-0049/208/1/9}{\color{magenta}\apjs}, \href{https://ui.adsabs.harvard.edu/abs/2013ApJS..208....9P}{208, 9}

\bibitem[{{Perryman} \& {ESA}(1997)}]{1997ESASP1200.....E}
{Perryman}, M.~A.~C. \& {ESA}, eds. 1997, \href{https://ui.adsabs.harvard.edu/abs/1997ESASP1200.....E}{ESA Special Publication, Vol. 1200, {The HIPPARCOS and TYCHO catalogues. Astrometric and photometric star catalogues derived from the ESA HIPPARCOS Space Astrometry Mission}}

\bibitem[{{Plaut}(1946)}]{1946BAN....10..153P}
{Plaut}, L. 1946, \bain, \href{https://ui.adsabs.harvard.edu/abs/1946BAN....10..153P}{10, 153}

\bibitem[{{Pojmanski}(2002)}]{2002AcA....52..397P}
{Pojmanski}, G. 2002, \href{http://dx.doi.org/10.48550/arXiv.astro-ph/0210283}{\color{magenta}\actaa}, \href{https://ui.adsabs.harvard.edu/abs/2002AcA....52..397P}{52, 397}

\bibitem[{{Pr{\v{s}}a}(2018)}]{2018maeb.book.....P}
{Pr{\v{s}}a}, A. 2018, {Modeling and Analysis of Eclipsing Binary Stars; The theory and design principles of PHOEBE}

\bibitem[{{Pr{\v{s}}a} {et~al.}(2022){Pr{\v{s}}a}, {Kochoska}, {Conroy}, {Eisner}, {Hey}, {IJspeert}, {Kruse}, {Fleming}, {Johnston}, {Kristiansen}, {LaCourse}, {Mortensen}, {Pepper}, {Stassun}, {Torres}, {Abdul-Masih}, {Chakraborty}, {Gagliano}, {Guo}, {Hambleton}, {Hong}, {Jacobs}, {Jones}, {Kostov}, {Lee}, {Omohundro}, {Orosz}, {Page}, {Powell}, {Rappaport}, {Reed}, {Schnittman}, {Schwengeler}, {Shporer}, {Terentev}, {Vanderburg}, {Welsh}, {Caldwell}, {Doty}, {Jenkins}, {Latham}, {Ricker}, {Seager}, {Schlieder}, {Shiao}, {Vanderspek}, \& {Winn}}]{2022ApJS..258...16P}
{Pr{\v{s}}a}, A., {Kochoska}, A., {Conroy}, K.~E., {et~al.} 2022, \href{http://dx.doi.org/10.3847/1538-4365/ac324a}{\color{magenta}\apjs}, \href{https://ui.adsabs.harvard.edu/abs/2022ApJS..258...16P}{258, 16}

\bibitem[{{Pr{\v{s}}a} \& {Zwitter}(2005)}]{2005ApJ...628..426P}
{Pr{\v{s}}a}, A. \& {Zwitter}, T. 2005, \href{http://dx.doi.org/10.1086/430591}{\color{magenta}\apj}, \href{https://ui.adsabs.harvard.edu/abs/2005ApJ...628..426P}{628, 426}

\bibitem[{{Ricker} {et~al.}(2015){Ricker}, {Winn}, {Vanderspek}, {Latham}, {Bakos}, {Bean}, {Berta-Thompson}, {Brown}, {Buchhave}, {Butler}, {Butler}, {Chaplin}, {Charbonneau}, {Christensen-Dalsgaard}, {Clampin}, {Deming}, {Doty}, {De Lee}, {Dressing}, {Dunham}, {Endl}, {Fressin}, {Ge}, {Henning}, {Holman}, {Howard}, {Ida}, {Jenkins}, {Jernigan}, {Johnson}, {Kaltenegger}, {Kawai}, {Kjeldsen}, {Laughlin}, {Levine}, {Lin}, {Lissauer}, {MacQueen}, {Marcy}, {McCullough}, {Morton}, {Narita}, {Paegert}, {Palle}, {Pepe}, {Pepper}, {Quirrenbach}, {Rinehart}, {Sasselov}, {Sato}, {Seager}, {Sozzetti}, {Stassun}, {Sullivan}, {Szentgyorgyi}, {Torres}, {Udry}, \& {Villasenor}}]{2015JATIS...1a4003R}
{Ricker}, G.~R., {Winn}, J.~N., {Vanderspek}, R., {et~al.} 2015, \href{http://dx.doi.org/10.1117/1.JATIS.1.1.014003}{\color{magenta}Journal of Astronomical Telescopes, Instruments, and Systems}, \href{https://ui.adsabs.harvard.edu/abs/2015JATIS...1a4003R}{1, 014003}

\bibitem[{{Savedoff}(1951)}]{1951AJ.....56....1S}
{Savedoff}, M.~P. 1951, \href{http://dx.doi.org/10.1086/106479}{\color{magenta}\aj}, \href{https://ui.adsabs.harvard.edu/abs/1951AJ.....56....1S}{56, 1}

\bibitem[{{Shi} {et~al.}(2022){Shi}, {Qian}, \& {Li}}]{2022ApJS..259...50S}
{Shi}, X.-d., {Qian}, S.-b., \& {Li}, L.-J. 2022, \href{http://dx.doi.org/10.3847/1538-4365/ac59b9}{\color{magenta}\apjs}, \href{https://ui.adsabs.harvard.edu/abs/2022ApJS..259...50S}{259, 50}

\bibitem[{{Soderhjelm}(1975)}]{1975A&AS...22..263S}
{Soderhjelm}, S. 1975, \aaps, \href{https://ui.adsabs.harvard.edu/abs/1975A&AS...22..263S}{22, 263}

\bibitem[{{Stassun} {et~al.}(2019){Stassun}, {Oelkers}, {Paegert}, {Torres}, {Pepper}, {De Lee}, {Collins}, {Latham}, {Muirhead}, {Chittidi}, {Rojas-Ayala}, {Fleming}, {Rose}, {Tenenbaum}, {Ting}, {Kane}, {Barclay}, {Bean}, {Brassuer}, {Charbonneau}, {Ge}, {Lissauer}, {Mann}, {McLean}, {Mullally}, {Narita}, {Plavchan}, {Ricker}, {Sasselov}, {Seager}, {Sharma}, {Shiao}, {Sozzetti}, {Stello}, {Vanderspek}, {Wallace}, \& {Winn}}]{2019AJ....158..138S}
{Stassun}, K.~G., {Oelkers}, R.~J., {Paegert}, M., {et~al.} 2019, \href{http://dx.doi.org/10.3847/1538-3881/ab3467}{\color{magenta}\aj}, \href{https://ui.adsabs.harvard.edu/abs/2019AJ....158..138S}{158, 138}

\bibitem[{{Strohmeier} {et~al.}(1964){Strohmeier}, {Knigge}, \& {Ott}}]{1964IBVS...66....1S}
{Strohmeier}, W., {Knigge}, R., \& {Ott}, H. 1964, Information Bulletin on Variable Stars, \href{https://ui.adsabs.harvard.edu/abs/1964IBVS...66....1S}{66, 1}

\bibitem[{{Torres} {et~al.}(2010){Torres}, {Andersen}, \& {Gim{\'e}nez}}]{2010A&ARv..18...67T}
{Torres}, G., {Andersen}, J., \& {Gim{\'e}nez}, A. 2010, \href{http://dx.doi.org/10.1007/s00159-009-0025-1}{\color{magenta}\aapr}, \href{https://ui.adsabs.harvard.edu/abs/2010A&ARv..18...67T}{18, 67}

\bibitem[{{van Hamme}(1993)}]{1993AJ....106.2096V}
{van Hamme}, W. 1993, \href{http://dx.doi.org/10.1086/116788}{\color{magenta}\aj}, \href{https://ui.adsabs.harvard.edu/abs/1993AJ....106.2096V}{106, 2096}

\bibitem[{{Wilson} \& {Devinney}(1971)}]{1971ApJ...166..605W}
{Wilson}, R.~E. \& {Devinney}, E.~J. 1971, \href{http://dx.doi.org/10.1086/150986}{\color{magenta}\apj}, \href{https://ui.adsabs.harvard.edu/abs/1971ApJ...166..605W}{166, 605}

\bibitem[{{Wolf} {et~al.}(2019){Wolf}, {Zasche}, {Ku{\v{c}}{\'a}kov{\'a}}, {Ma{\v{s}}ek}, {Ho{\v{n}}kov{\'a}}, {Jury{\v{s}}ek}, {Paschke}, {{\v{S}}melcer}, \& {Zejda}}]{2019AcA....69...63W}
{Wolf}, M., {Zasche}, P., {Ku{\v{c}}{\'a}kov{\'a}}, H., {et~al.} 2019, \href{http://dx.doi.org/10.32023/0001-5237/69.1.5}{\color{magenta}\actaa}, \href{https://ui.adsabs.harvard.edu/abs/2019AcA....69...63W}{69, 63}

\bibitem[{{Wolf} \& {Zejda}(2005)}]{2005AA...437..545W}
{Wolf}, M. \& {Zejda}, M. 2005, \href{http://dx.doi.org/10.1051/0004-6361:20041868}{\color{magenta}\aap}, \href{https://ui.adsabs.harvard.edu/abs/2005A&A...437..545W}{437, 545}

\bibitem[{{Wolf} {et~al.}(2008){Wolf}, {Zejda}, \& {de Villiers}}]{2008MNRAS.388.1836W}
{Wolf}, M., {Zejda}, M., \& {de Villiers}, S.~N. 2008, \href{http://dx.doi.org/10.1111/j.1365-2966.2008.13527.x}{\color{magenta}\mnras}, \href{https://ui.adsabs.harvard.edu/abs/2008MNRAS.388.1836W}{388, 1836}

\bibitem[{{Wolf} {et~al.}(2022){Wolf}, {Zejda}, {Ma{\v{s}}ek}, {Ku{\v{c}}{\'a}kov{\'a}}, {de Joode}, {Uhla{\v{r}}}, \& {Zasche}}]{2022NewA...9201708W}
{Wolf}, M., {Zejda}, M., {Ma{\v{s}}ek}, M., {et~al.} 2022, \href{http://dx.doi.org/10.1016/j.newast.2021.101708}{\color{magenta}\na}, \href{https://ui.adsabs.harvard.edu/abs/2022NewA...9201708W}{92, 101708}

\bibitem[{{Zasche} {et~al.}(2014){Zasche}, {Wolf}, {Vra{\v{s}}til}, {Li{\v{s}}ka}, {Skarka}, \& {Zejda}}]{2014A&A...572A..71Z}
{Zasche}, P., {Wolf}, M., {Vra{\v{s}}til}, J., {et~al.} 2014, \href{http://dx.doi.org/10.1051/0004-6361/201424273}{\color{magenta}\aap}, \href{https://ui.adsabs.harvard.edu/abs/2014A&A...572A..71Z}{572, A71}

\end{thebibliography}

\end{document}